\documentclass[a4paper,fleqn]{cas-dc}

\usepackage[numbers]{natbib}
\usepackage{pgfplots}
\pgfplotsset{compat=1.17}
\usepackage{graphicx}
\usepackage[dvipsnames]{xcolor}
\usepackage{makecell}
\usepackage{tikz}
\usepackage[ruled,linesnumbered]{algorithm2e}
\usepackage{graphicx,listings,tikz}
\usepackage{pgfplots}
\usepackage{etoolbox}
\usepackage{breakcites}
\usepackage{wrapfig}
\usepackage{makecell}
\usepackage{caption}
\usepackage{hyperref}
\usepackage{subcaption}
\usepackage{calc}
\usepackage{xfp}
\usepackage{float}
\usepackage{amsmath,amsthm,amssymb}
\usepackage{booktabs}
\usepackage{bm}
\theoremstyle{definition}
\newtheorem{definition}{Definition}
\newtheorem{example}{Example}

\newcommand{\col}{\fpeval{\x *100 /35}}
\newcommand{\toolname}{ADAPT\xspace}
\newcommand{\coderepo}{https://gitfront.io/r/anonymous-submitter/P2LRhxvh9L7z/ADAPT/}

\usetikzlibrary{shapes.geometric,arrows,fit,calc,positioning,automata,backgrounds,fit,patterns}

\usepackage{tablefootnote}
\usepackage[section]{placeins}

\usetikzlibrary{shapes,shapes.geometric,arrows,fit,calc,positioning,automata}
\tikzset{eli state/.style={draw,ellipse}}
\tikzset{rect state/.style={draw,rectangle}}
\tikzset{diamond state/.style={draw,diamond}}
\tikzset{blank state/.style={draw=none}}
\newcounter{barcount}
\tikzset{barchart/.cd,y distance/.initial=3em,
bar height/.initial=2em,width/.initial=6cm,
bar text/.style={font=\sffamily,text depth=0.25em},
description/.style={font=\sffamily,text depth=0.25em},
colors/.initial={"orange!80","blue!40","red","green!70!black"}}

\def\addlegendimage{\csname pgfplots@addlegendimage\endcsname}

\newcommand{\PStates}{\ensuremath{\mathcal{P}}\xspace}
\newcommand{\Attacks}{\ensuremath{\mathcal{A}}\xspace}

\newcommand{\xmark}{%
\tikz[scale=0.23] {
    \draw[line width=0.7,line cap=round] (0,0) to [bend left=6] (1,1);
    \draw[line width=0.7,line cap=round] (0.2,0.95) to [bend right=3] (0.8,0.05);
}}
\newcommand{\cmark}{%
\tikz[scale=0.23] {
    \draw[line width=0.7,line cap=round] (0.25,0) to [bend left=10] (1,1);
    \draw[line width=0.8,line cap=round] (0,0.35) to [bend right=1] (0.23,0);
}}

\newcommand\score[2]{%
  \pgfmathsetmacro\pgfxa{#1 + 1}%
  \tikzstyle{scorestars}=[star, star points=5, star point ratio=2.25, draw, inner sep=0.15em, anchor=outer point 3]%
  \begin{tikzpicture}[baseline]
    \foreach \i in {1, ..., #2} {
      \pgfmathparse{\i<=#1 ? "black" : "white"}
      \edef\starcolor{\pgfmathresult}
      \draw (\i*1em, 0) node[name=star\i, scorestars, fill=\starcolor]  {};
    }
    \pgfmathparse{#1>int(#1) ? int(#1+1) : 0}
    \let\partstar=\pgfmathresult
    \ifnum\partstar>0
      \pgfmathsetmacro\starpart{#1-(int(#1)}
      \path [clip] ($(star\partstar.outer point 3)!(star\partstar.outer point 2)!(star\partstar.outer point 4)$) rectangle 
      ($(star\partstar.outer point 2 |- star\partstar.outer point 1)!\starpart!(star\partstar.outer point 1 -| star\partstar.outer point 5)$);
      \fill (\partstar*1em, 0) node[scorestars, fill=black]  {};
    \fi
  \end{tikzpicture}%
}

\pgfplotsset{ 	ybar/.append style={ area legend} }
\def\addlegendimage{\csname pgfplots@addlegendimage\endcsname}

\def\tsc#1{\csdef{#1}{\textsc{\lowercase{#1}}\xspace}}
\tsc{WGM}
\tsc{QE}
\tsc{EP}
\tsc{PMS}
\tsc{BEC}
\tsc{DE}

\begin{document}
\let\WriteBookmarks\relax
\def\floatpagepagefraction{1}
\def\textpagefraction{.001}
\shorttitle{Automated Penetration Testing: Formalization and Realization}
\shortauthors{C. Skandylas, M. Asplund}

\title [mode = title]{Automated Penetration Testing: Formalization and Realization}

\author[1]{Charilaos Skandylas}[orcid=0000-0001-0000-0000]
\cormark[1]
\ead{charilaos.skandylas@liu.se}


\affiliation[1]{organization={Department of Computer and Information Science, Linköping University},
                city={Linköping},
                country={Sweden}}

\author[1]{Mikael Asplund}[orcid=0000-0003-1916-3398]
\ead{mikael.asplund@liu.se}


\begin{abstract}
Recent changes in standards and regulations, driven by the increasing importance of software systems in meeting societal needs, mandate increased security testing of software systems. Penetration testing has been shown to be a reliable method to asses software system security. However, manual penetration testing is labor-intensive and requires highly skilled practitioners. Given the shortage of cybersecurity experts and current societal needs, increasing the degree of automation involved in penetration testing can aid in fulfilling the demands for increased security testing.
In this work, we formally express the penetration testing problem at the architectural level and suggest a general self-organizing architecture that can be instantiated to automate penetration testing of real systems. We further describe and implement a specialization of the architecture in the ADAPT tool, targeting systems composed of hosts and services. 
We evaluate and demonstrate the feasibility of ADAPT by automatically performing penetration tests with success against: Metasploitable2, Metasploitable3, and a realistic virtual network used as a lab environment for penetration tester training.
\end{abstract}


\begin{keywords}
Automated Penetration Testing, \sep  Attack Planning, \sep Attack Automation, \sep Formal Model
\end{keywords}

\maketitle

\section{Introduction}

Computer systems are ubiquitous in society and their security is becoming a primary concern not just for businesses and individuals, but also for national security.
Critical infrastructure, healthcare services, and governmental services are examples of areas where cybersecurity is becoming a top priority.
In the last few years, security concerns have led to an increase in regulations and industry requirements on rigorous security engineering practices being applied for such systems.
Multiple regulations and standards (e.g., HIPA\footnote{\url{csrc.nist.gov/pubs/sp/800/66/r2/final}}, PCI DCSS\footnote{\url{east.pcisecuritystandards.org/document_library?category=pcidss\&document=pci_dss}} and SWIFT CSF\footnote{\url{www2.swift.com/knowledgecentre/publications/cscf_dd/53.0}}) nowadays mandate the application of security testing, i.e., penetration testing.

On one hand, penetration testing is just another form of testing which is a fundamental technique in all of engineering and it inherits much of its principles from software testing in particular.
On the other hand, penetration testing is particular in the sense that it overlaps considerably with practices and methods that are performed by criminals and unfriendly nation agencies.
As a result, a complete ecosystem of tools and methods that require a very particular form of expertise to deploy and operate properly has emerged.
Despite increasing efforts to train students in ethical hacking, there is an enormous lack of human resources that can effectively perform penetration
testing at the level that we need to ensure that future systems will be more secure and able to withstand increasingly advanced and frequent attacks.


A potential solution to this problem is to provide better tools that automate a larger part of the penetration testing process,
thereby reducing the human effort and training required to start deploying penetration testing on systems.  
There are a number of proposals in the literature on automated penetration testing. 
One direction which has been pursued is to use attack graphs to generate plans for the attacker~\cite{DBLP:journals/corr/ObesSR13}.
This can potentially allow finding the best path through a system, however, there is a lack of works that demonstrate how such plans can be transferred to real penetration testing scenarios.
Another approach that has been explored by several works~\cite{9870951,10335188} is to use machine learning to train a model to perform attacks based on available data.
The downsides this approach are the lack of available data to train, and the need for retraining once enough system changes occur.
Finally, reinforcement learning(RL) approaches~\cite{Zhou2019,9229752,app11198823} allow agents to operate without an initial training phase, however, the instantiation to real systems has yet to be   widely demonstrated yet.
Moreover, what is meant by the term automated penetration testing is also interpreted differently in different papers,
with the main focus of most works having been attack planning, which is just one part of the penetration testing process.
This hints to a need for a precise definition of which elements are required to automate the penetration testing process when targeting realistic systems.
Most existing works have evaluated their effectiveness in network simulator environments,
rather than real networks which may exhibit different behaviors when it comes to the effectiveness and results of exploitation.
To our knowledge, there exist no automated penetration testing approaches that can perform the whole penetration testing process,
i.e., that is able to perform attack planing and runtime decision making to handle changes in the system state or newly discovered information
in addition to providing the required attack automation capabilities to target real-world systems.

In this paper, we formally express the penetration testing problem at the architectural level with its constituent elements being components and interfaces.
This high-level representation can be instantiated on a number of different actual systems such as a network of hosts, or a single host with multiple processes, component-based or service-oriented systems, etc.
By precisely formulating the automated penetration testing problem, we make it easier to reason about different performance metrics,
as well as for the community to extend and refine the formulation to allow for alternative objectives and/or constraints.
Our proposed formulation is more general than corresponding RL or AI-planning based formulations as it is possible to easily construct a mapping from most RL-based or AI planning-based formulations to our formalism.

Moreover, we present an architecture for solving the automated penetration testing problem based on autonomic computing principles (i.e., using a MAPE-K loop~\cite{Kephart:2003:VAC:642194.642200}).
The architecture is able to handle system dynamics and make decisions at runtime to provide efficiency, e.g., not repeating attacks with the same/similar goals.
It can also make decisions at runtime to provide effectiveness by dynamically selecting exploitation tools and deploying post-exploitation tools on the exploited part of the system to discover
as much of the system as possible and exploit as deeply as possible.

Finally, we provide \toolname, a concrete implementation of this architecture that utilizes state-of-the-art penetration testing tools such as metasploit, sqlmap, etc.

\toolname is a black-box penetration testing tool that also provides white-box and gray-box capabilities.
We evaluate our concrete implementation and demonstrate that our approach is practical in a realistic setting.
Given a sufficient set of penetration testing tools, the operation of which it automates, \toolname successfully performs penetration testing, without human interaction, against
the Metasploitable2 and Metasploitable3 VMs, as well as against a realistic virtual network.

To summarize, the contributions of this paper are:
\begin{itemize}
    \item Formulating penetration testing in an architectural context, providing a formal problem description that is expressive enough to incorporate previous formulations in a unified framework.
    \item Proposing a generic automated penetration testing architecture, employing an autonomic manager to reason about the general, high-level decisions and to provide attack automation capabilities.
    \item Implementing and evaluating the architecture against a realistic network setup.
\end{itemize}

The rest of this paper is structured as follows.
In Section~\ref{sec::related.work}, we present related work.
Section~\ref{sec::preliminares} provides a high-level introduction to architectural modeling for security and to self-adaptive systems.
Section~\ref{sec:problem} introduces and formulates the penetration testing problem. Section~\ref{sec::architecture} discusses our general architecture and our instantiation for systems composed of hosts and services.   Section~\ref{sec::evaluation} details the evaluation of two case studies.
The paper conclusion is found in Section \ref{sec::conclusion}. 

\section{Related Work}
\label{sec::related.work}

\begin{table*}[ht]
  \caption{Related Work Comparison}
  \label{tab:related.work.comparison}
  \small
  \centering
    \let\Tabular\tabular
    \def\tabular{\Tabular}    
    \scalebox{0.8}{
    \begin{tabular}{c|ccc|ccc|c|c|c}
    \hline
    \textbf{Reference(s)} & \multicolumn{3}{c|}{\textbf{Pentest Type}} & \multicolumn{3}{c|}{\textbf{Features}} & \makecell{\textbf{Evaluation} \\ \textbf{Environment}} & \makecell{\textbf{Solution} \\ \textbf{Domain}} & \makecell{Autonomy \\ Level} \\
    \hline
      & \textbf{White-box} & \textbf{Gray-box} & \textbf{Black-box}  &\makecell{\textbf{Attack} \\ \textbf{Automation}}  &\makecell{\textbf{Attack} \\ \textbf{Planning}} & \makecell{\textbf{Runtime} \\ \textbf{Decision-making}} &  & \\
    \hline %
    \cite{DBLP:journals/corr/ObesSR13,Sarraute_2011,Sarraute2013PenetrationT} & \cmark & \xmark & \xmark & \score{1}{1} & \cmark & \xmark & \makecell{Simulation } & \makecell{AI \\ Planning} & 3 \\
    \hline
    \cite{9394285} & \cmark & \xmark & \xmark & \score{0.5}{1} & \cmark & \xmark & \makecell{ Simulation } & \makecell{Reinforcement \\ Learning} & 3 \\
    \hline
    \cite{Zhou2019} & \xmark & \xmark & \cmark & \score{0.5}{1} & \cmark & \xmark & \makecell{VM Network} & \makecell{Reinforcement \\ Learning} & 3 \\
    \hline
    \cite{9229752} & \cmark & \xmark & \xmark & \score{0.0}{1} & \cmark & \xmark & \makecell{Simulation } & \makecell{Reinforcement \\ Learning} & - \\
    \hline
    \cite{CHEN2023103055} & \cmark & \xmark & \xmark & \score{0.5}{1} & \cmark & \xmark & \makecell{Simulation } & \makecell{Reinforcement \\ Learning} & - \\
    \hline
    \cite{app11198823,LI2023103358} & \xmark & \cmark & \xmark & \score{0.0}{1} & \xmark & \xmark & \makecell{Simulation } & \makecell{Reinforcement \\ Learning} & - \\
    \hline
    This work & \cmark & \cmark & \cmark & \score{1}{1}  &  \cmark & \cmark & \makecell {VM Network} & \makecell{Autonomic \\ Computing} & 3 \\
    \hline    
    \end{tabular}
    }
    \\ \score{1}{1} Modular/extensible automation capabilities,
    \score{0.5}{1} Wrapper over Metasploit, \score{0.0}{1} Penetration Test module not implemented
\end{table*}

The main body of work related to automating penetration testing revolves around automating attack selection and planning. There are three main directions that have been followed: (i) graphical or probabilistic models, 
(ii) deep learning and (iii) reinforcement learning.

Attack automation can be partially provided in varying degrees by off-the-shelf tools including metasploit~\cite{7028682},
cobalt-strike~\footnote{https://www.cobaltstrike.com/product} and canvas~\footnote{https://www.immunityinc.com/products/canvas/},
however, these tools, do not provide any reasoning or automated decision making capabilities required to evaluate the security status and decide what the next target or attack is.  
Moreover, attack automation has been demonstrated in the DARPA Grand Cyber Challenge~\cite{Lee2015}.
Both off-the-shelf tools and the capabilities demonstrated by the DARPA Grand Cyber Challenge reside
at the 2nd level of autonomy with reference to the NATO IAG pre-feasibility study on autonomous operations~\cite{NATOAutonomous}.
Automated, at runtime decision making and reasoning are essential to move past level 2 to higher autonomy
levels. In the context of automating penetration testing against realistic systems,
runtime reasoning and decision-making are essential to deal with changing system dynamics,
such as attacks failing due to changes or unreliability in the environment(network) or the target
system itself, as well as to adapt at runtime to handle possible mitigations that might be deployed or
exploit newly discovered avenues of exploitation.
Attack automation is an important part of penetration testing, however, on its own cannot provide the level of autonomy required to automate penetration testing in its entirety.
Our work aims at level 3 or higher autonomy.

When it comes to attack planning,
Obes et. al.~\cite{DBLP:journals/corr/ObesSR13} utilize attack graphs to generate a planning domain definition language (PDDL) representation of the attack model and generate an attack plan for penetration testing tools. Sarraute et.al.~\cite{Sarraute2013PenetrationT} encode the penetration testing problem as a partially observable Markov decision process (POMDP). The authors use empirical evaluation on a pair of machines to validate the solution and capture uncertainty in the results of attack action using PDDL frameworks that include probabilistic effects. Sarraute et.al~\cite{Sarraute_2011} also 
provide a scenario specific probability-aware approach for scalable attack planning.

Chowdhary et. al.~\cite{9394285} propose the ASAP framework that identiﬁes the dependencies between vulnerabilities, and network connectivity to provide domain-speciﬁc reward modelling. The transition probabilities are associated with a network structure, and access complexity and the reward values are associated with CVSS scores of vulnerabilities. They utilize a Reinforcement Learning (RL) framework based on Deep Q-Networks to learn efficient pentest plans on a large-scale network.  Zhou et. al.~\cite{Zhou2019} introduce an action decomposition scheme which is efficient and versatile in developing an attacking strategy in different network configurations. The proposed method utilizes multi-agent RL to decompose the action space into smaller sets and train individual RL agents in each of the action subsets. 

Hu et. al.~\cite{9229752} present an automated penetration testing framework to find the best attack path for a given topology, operating similarly to a human attacker. They construct an attack representation matrix by using MulVAL and Depth-First Search (DFS) algorithms. A Deep Q-Learning network method is used to analyse the attack matrix and find the optimal attack path.
In~\cite{app11198823}, the authors model penetration testing as a Markov decision process (MDP) problem.  They present a deep q-network to train a penetration testing agent that can learn the optimal attack policy without prior knowledge. Faillon et. al.~\cite{10.1007/978-3-031-70903-6_16}, provide a policy-based approach to improve the generalization capabilities of a hierarchical reinforcement learning attack agent to previously unknown scenarios.
Li et. al.~\cite{LI2023103358}, propose a hierarchical deep learning method that takes prior expert knowledge  into consideration to to guide exploration and reduce the state and action spaces.
Chen et.al~\cite{CHEN2023103055}, exploit generative adversarial imitation learning in concert with expert knowledge
to improve convergence when deciding the next actions to be taken by a penetration testing agent in large action spaces.
The feasibility of the above approaches has been demonstrated by utilizing network simulators the most common being NASim and CybORG~\cite{Standen2021CybORGAG}. The only approach, other than our own, that targets realistic systems and performs black-box penetration testing, albeit with limited attack automation capabilities, is ~\cite{Zhou2019}. Unfortunately, the authors do not provide their code or an environment replication package for the experiments, making it difficult to compare against that approach quantitatively.

In contrast to approaches that aim to automate attack planning, our approach provides a general end-to-end architecture to automate the whole penetration testing process.
Our reference implementation, \toolname provides both attack planning and attack automation capabilities and thus can be used to target real systems as shown by our VM case study.
Moreover, most approaches based on planning, deep or reinforcement learning, require varying amounts information to be known about the target system either to train their learning models
or to generate the planning strategy. Our approach requires no knowledge of the system architecture, vulnerabilities or functionality.
At the same time, our approach allows  integrating existing methods for attack planning in the analysis and planning phases, which is an avenue for future work.

Recently, a number of works employing LLMs to automate penetration testing have been proposed~\cite{299699,Hilario2024}, demonstrating initial success on single VM case studies.
Our concrete implementation  outperforms them in
the number of services exploited, running time and scalability.

We summarize the related work and compare with our approach in table~\ref{tab:related.work.comparison}.
We classify the pentest type as white-box, gray-box and black-box based on the information about the target system and its vulnerabilities required to train the models or perform the penetration test itself.

\section{Preliminaries}
\label{sec::preliminares}
This section covers the required prerequisites for this work including  architectural modeling for security, self-adaptive systems and utility-based decision making.

\subsection{Architectural Modeling for Security}
We define a software architecture as a composition of (i) components, (ii) interactions and (iii) architectural properties.
{Components specify the behavior of individual parts of the
system. Each component comprises a set of interfaces that can be invoked 
to provide the system functionality associated with them. Each interface
is associated with a set of capabilities. A capability is the ability to perform a certain set of actions on a component or its environment. An interaction is an invocation of one or more target interfaces from a source interface to realize the corresponding functionality.
An architectural property represents an important quality about a component or an interface.
A security-informed architecture is a software architecture whose properties concern security-related information including the vulnerabilities associated with each component, the interfaces that can be invoked to exploit those vulnerabilities and how vulnerabilities can be combined to form more complex attacks that allow the attacker to compromise the system. We consider a vulnerability~\cite{236681} to be: "An error, flaw, or mistake in computer software that permits or causes unintended
behavior to occur".
Borrowing the weird machines~\cite{8226852,BratusShubina2017} conceptual definition and  transitioning the concept to security-informed architectures,
we consider an exploit to be: "crafted Input" to a \emph{vulnerable interface} that causes adversary controlled unintended behavior".

\begin{definition}
We define a security-informed architecture $\mathcal{S}$ as a pair of sets $\langle \mathcal{C}, \mathcal{R} \rangle$, where:
\begin{itemize}
    \item a component $c \in \mathcal{C}$ is defined as a tuple $ \langle \mathcal{I}_{c}, \Pi_{c} \rangle$ where:  
    $\mathcal{I}_c$ is the set of $c$'s interfaces used for communication with other components. 
    An interface $i \in \mathcal{I}_{c}$ is a tuple $\langle \mathcal{K}_{i}, V_{i}, \Pi_{i} \rangle$ where: $\mathcal{K}_{i}$ refers to the set capabilities, $V_{i}$ refers to the possibly empty set of vulnerabilities, and $\Pi_{i}$ refers to a set of security-relevant architectural properties of $i$.
    $\Pi_{c}$ refers to a set of security relevant architectural properties of $c$.
     \item $\mathcal{R}$ is a set of component interactions. An interaction $r$ is a pair $(i_{s}, i_{t})$ where $i_{s}$ and $i_{t}$ are the source and target interfaces of the interaction.
\end{itemize}
\end{definition}
We assume that the sets of interfaces of each component are disjoint, hence,
we define: $\mathcal{I} = \bigcup\limits_{c \in \mathcal{C}}\mathcal{I}_{c}$ to be the set of all interfaces in the system. Similarly, the set of each interface's capabilities are assumed to be disjoint, thus, we define $\mathcal{K}_{\mathcal{I}} = \bigcup\limits_{i \in \mathcal{I}}\mathcal{K}_{i}$ to be the set of all capabilities of all interfaces in the system, where $\mathcal{K}_{\mathcal{I}} \subseteq \mathcal{K}$ the set of all system capabilities.

In the context of a security-informed architecture, the capabilities $\mathcal{K}_{i}$ of an interface $i$, can be partitioned into $\mathcal{K}_{i}^{I}$, the set of intended capabilities (by the system designer and implementer), and $\mathcal{K}_{i}^{U}$, the set of unintended capabilities. The set of unintended capabilities can be further partitioned into: (i) the set of  non-controllable capabilities $\mathcal{K}_{i}^{NC}$ and the set of adversary controllable (weird) capabilities $\mathcal{K}_{i}^{W}$.

In other words, when an interface $i$ is invoked, its behavior is drawn from the set of capabilities associated with that interface $\mathcal{K}_{i}$. There are three possible cases for the interface behavior:
\begin{itemize}
    \item If the input is valid (as expected by the system designer), then the interface behaves as intended, i.e., it draws its functionality from $\mathcal{K}_{i}^{I}$.
    \item If the input is invalid (as expected by the system designer), but leads the interface to adversary-controllable unintended behavior, i.e., exploitation, the interface draws its functionality from the set of adversary-controllable capabilities, $\mathcal{K}_{i}^{W}$, 
    \item If the input is invalid and non-controllable,
    the interface behavior is not useful to an adversary,
    e.g., it might lead to a crash or to an error state. 
    The interface draws its functionality from the set of non-controllable capabilities, $\mathcal{K}_{i}^{NC}$.
\end{itemize}

\subsection*{Running Example}
Figure~\ref{fig:running.example} shows an example security-informed architecture. It comprises 5 components: \emph{WebServer}, \emph{APIGate}, \emph{Authentication}, \emph{Business Logic} and \emph{Database}.
  Of those, the Webserver and APIGate components are exposed to the internet, and all components are
  connected in a private virtual LAN network. 
  The system models a small-scale web architecture.
The web server acts as the
  presentation layer, while the APIGate combined with the Authentication and Business Logic microservices provide the application layer, the storage layer needs are fulfilled by the database.
  Each components has a number of vulnerable interfaces that are also shown in the figure. 

\begin{figure}[ht]
  \centering
  \includegraphics[scale=0.15]{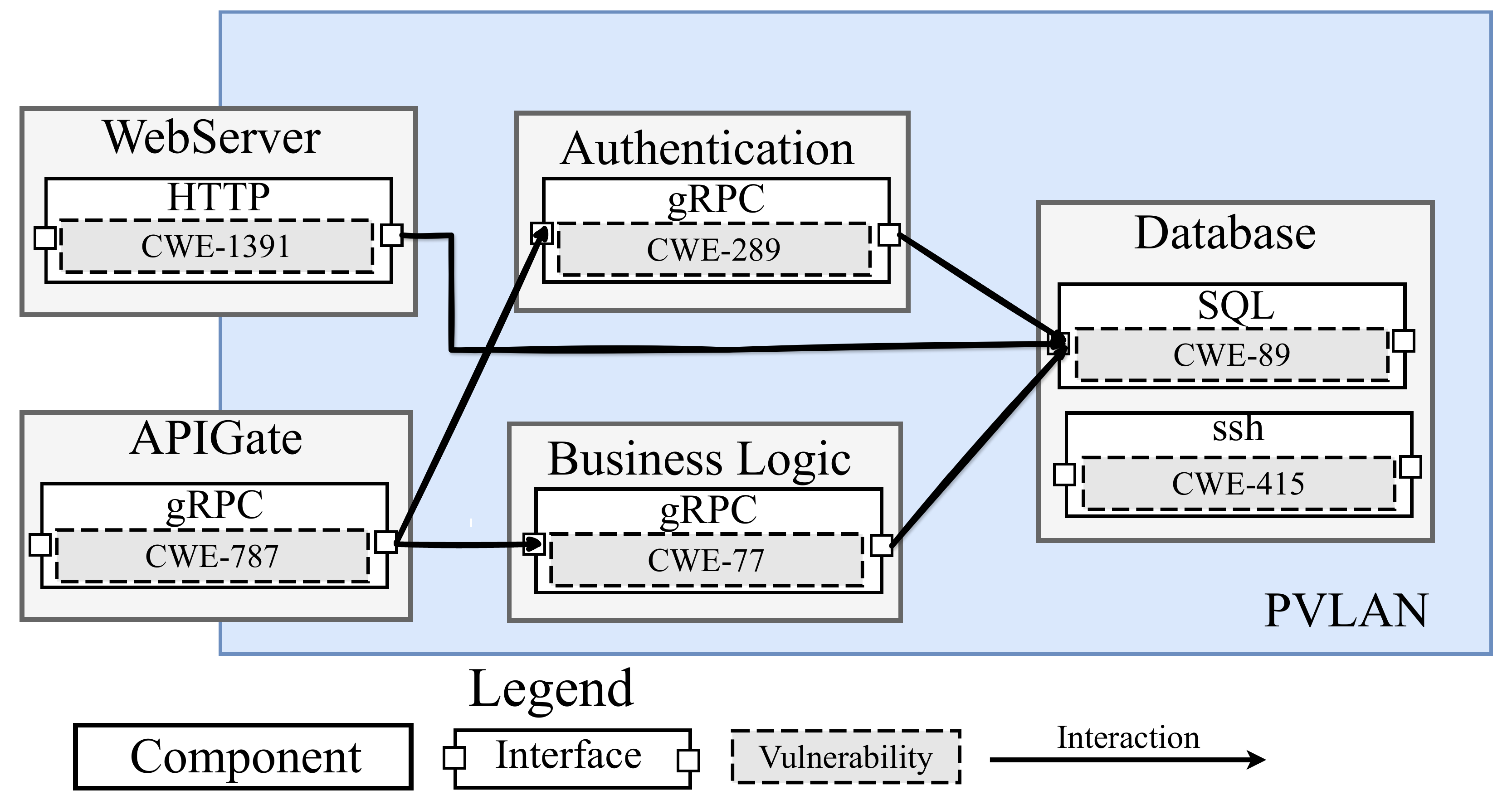}
  \caption[justification=centering]{An Example security-informed Architecture.}
  \label{fig:running.example}
\end{figure}

\subsection{Self-Adaptive Systems}
A self-adaptive system is a software system that can dynamically and autonomously alter its architecture or behavior at run-time to more effectively achieve stakeholder goals and extra-functional qualities~\cite{Cheng2009}. A self-adaptive system might additionally adapt itself to accommodate changes in its context or its environment.
A self-adaptive system architecture often comprises three basic
elements~\cite{Weyns2019}: (i) the environment, (ii) the managed system, and (iii) the
managing system. The managing system controls the managed system and contains the adaptation logic required to achieve adaption goals. The managing system monitors both the environment and the managed system and adapts the latter when required to progress towards its adaptation goals. 

The primary building block of a managing system is an autonomic manager~\cite{Kephart:2003:VAC:642194.642200}, that comprises four phases:
Monitoring, Analysis, Planning, and Execution. All four phases share a common Knowledge Base.
The Monitoring phase gathers data from the managed system and the environment, and updates the information stored in the Knowledge Base with the updated view of the world. The Analysis phase exploits the updated information stored in the Knowledge Base to determine whether an adaptation is required. When an adaptation is triggered, the Planning phase constructs a plan by combining a set of adaptation actions to meet the adaptation goals. The adaptation plan is executed by the Execution phase that performs the required changes to the managed system.

\subsection{Utility-based Decision Making}
In utility-based decision making~\cite{vonneumann.morgenstern47,anand1995foundations}, a decision is a choice of one among a set of options $o \in O$. A factor, $f \in F$ refers to a concern that influences the decision. The function $v_{f}(o)$ returns the value of an option $o$ with regards to a concern $c$. The function $u_{f}(o)$ returns corresponding the normalized value. Factors are ascribed different 
levels of importance given by a set of weights, that sum up to 1, i.e., $\sum\limits_{f \in F} w_{f} = 1$. The utility of an option $o$, given a set of factors $F$, is then defined as:
\begin{equation}
 U(o) = \sum\limits_{f \in F}^{} w_{f} \cdot u_{f}(o)     \label{eq:utility}
\end{equation}

\section{Problem Formulation}
\label{sec:problem}
We assume that the system to be penetration tested is modeled as a security-informed architecture that is initially unknown to the penetration tester.
We refer to that system and its architecture as the target system or target architecture in the rest of this paper.
We define penetration testing as the process by which the penetration tester attempts to discover as much of the target system architecture and acquire as many of its capabilities as possible.

Capabilities usually grant increased/additional privileges or provide means to non-intended functionality, e.g., by gaining remote code execution capabilities on a component, however, capabilities can also include actions that can lead to reducing the system service level, e.g.,
when the penetration tester gains denial or service or similar capabilities.
In this work, given our stated goal, the penetration tester will only utilize such capabilities if they can be used to further expand their knowledge of the system, e.g., learning about countermeasures or to further expand their gained capabilities.

The aim of this section is to provide a unified problem formulation that is independent of the solution domain and expressive enough to encompass the approaches we discussed in our related work work section, in addition to the generic architecture 
and concrete implementation that we propose in Section~\ref{sec::architecture}. 
The two main elements of our problem formulation in this section are a labeled transition system~\cite{GLABBEEK20013} (LTS) and a game strategy~\cite{kwiatkowska:LIPIcs.ICALP.2016.4}.
Intuitively, the LTS defines the action space while the strategy defines the solving approach. Table~\ref{tab:mapping} gives a comparison of the different approaches.

\begin{table}
  \caption{Automated Penetration Testing Approaches}
  \label{tab:mapping}
  \centering
  \small
    \let\Tabular\tabular
    \def\tabular{\Tabular}
    \scalebox{0.85}{
    \begin{tabular}{ccc}
      \toprule
      \textbf{Approach} &  \textbf{Transition System} & \textbf{Strategy Basis} \\
      \midrule
       AI Planning & \makecell{Labeled Transition \\ System$^{*}$}  & \makecell{Heuristic Search, \\ Temporal Planning, \\ POMDP Solving }   \\
      \midrule
       \makecell{Model-free \\ reinforcement learning} & \makecell{Finite State\\ Machine$^{**}$} & \makecell{Q-Learning, \\ Deep Q-Learning} \\
      \midrule
       This work & \makecell{Labeled Transition \\ System} & \makecell{Autonomic Computing, \\ Utility-based \\Decision Theory }   \\
     \bottomrule
    \end{tabular}
    }\\\scriptsize{A Partially Observable Markov Decision Process can be represented by a LTS that encodes observation, probabilities and rewards via labeling functions$^{*}$}\\
    \scriptsize{A Markov Decision Process with unknown rewards and probabilities is a finite state machine which can be represented as a restricted LTS$^{**}$}\\
\end{table}

In the rest of this section, we first formalize constituent elements required to formally express the penetration testing problem and then build on those definitions to specify penetration testing at the architectural level.


\subsection{Pentest State, Attacks, and Scans}
The penetration tester makes decisions based on her current knowledge of
the target system architecture and her current capabilities.
To achieve her goal, the penetration tester performs attacks and scans targeting the target system components.

\begin{definition}
\label{def:pentest.state}
Given a target system architecture, modelled as a security-informed architecture $\mathcal{S}= \langle \mathcal{C}, \mathcal{R} \rangle$, a penetration test state is a tuple $P = \langle K, C, I \rangle$ where:
\begin{itemize}
    \item $K \subseteq \mathcal{K}$, the set of the penetration tester's capabilities,
    \item $C \subseteq \mathcal{C}$, the set of components that the penetration tester has knowledge of,
    \item $I \subseteq \mathcal{I}$, the set of interfaces that the penetration tester has knowledge of
\end{itemize}
The penetration tester's capabilities refer to the actions that the penetration tester may take to affect the system, which can be both benign and malicious. We consider benign actions
  those that do not necessarily lead to component compromise and whose danger level depends on the context. Examples of those capabilities include: listing the running processes of a system, opening a new port or creating a new user. Conversely, malicious actions are those that expressly aim at system compromise. Examples of such actions include, exploiting a memory corruption vulnerability,
  running a password cracker, etc.
In this context, knowledge of a component refers to knowing that the component is part of the target system, while knowledge of an interface 
refers to knowing that the interface is part of the system in addition to
some additional information that can characterize the interface.
For instance, when the penetration tester initially finds out about the Database component in the running example, she only has knowledge about the SQL interface. However, once the SQL injection vulnerability is utilized to upload a remote shell, the penetration tester can list the running services on the Database component and learn about the vulnerable ssh interface also residing in the same component.
\end{definition}

The penetration tester can expand her capabilities by performing attacks, and her presently known components and interfaces by performing scans. {An attack refers to a sequence of attack steps that might involve the exploitation of vulnerabilities in a component's interface which results in the penetration tester gaining new capabilities. A scan corresponds to a series of actions that lead to the penetration tester expanding their knowledge
of the target system's components or interfaces.

\begin{definition}
\label{def:attack}
  Given a security-informed architecture $\mathcal{S} = \langle \mathcal{C},\mathcal{R} \rangle$ and a penetration test state $P = \langle K, C, I \rangle$, an attack $A$ is a sequence of attack steps. An attack step is a tuple $\alpha = \langle r, v, K_{0}, K_+ \rangle$ where:
\begin{itemize}
    \item $r = \langle s, d \rangle \in R$, is the interaction, bounded by the source and target interfaces $s$ and $d$, that invokes the functionality required to facilitate the attack step,
    \item $v \subseteq V_{d}$, is the (possibly empty) set of vulnerabilities,
    \item $K_{0} \subseteq K$, is the set of prerequisite capabilities that the penetration tester must have for the attack step $\alpha$ to be possible,
    \item $K_+ \subseteq \mathcal{K}_{I}$ is the set of capabilities gained by the penetration tester when the attack step succeeds.
\end{itemize}
  After an attack step succeeds, the penetration tester capabilities are updated as follows:  $K = K \cup K_+$. In that event, the penetration tester, in addition to gaining a subset of the capabilities provided by interface $d$,
  might gain capabilities not associated with the interface itself. For example, a penetration tester might get the capability to access a new host in a network by disabling a firewall rule.
We mark the set of capabilities gained by performing an attack $A$ with $K_{A}$, where $K_{A} = \bigcup\limits_{\alpha \in A}K_{+}$.
\end{definition}

\begin{definition}
\label{def:scan}
Given an architecture $\mathcal{S}$ and penetration test state $P$, a scan 
is a tuple $\sigma = \langle K_{0}, C_{+},I_{+} \rangle$ where:
\begin{itemize}
    \item $K_{0} \in K$, the set of prerequisite capabilities that the penetration tester must hold to perform the scan,
    \item $C_{+}$, the set of components identified by the scan, 
    \item $I_{+}$, the set of interfaces identified by the scan 
\end{itemize}
After a a scan's completion, the penetration test state is updated as follows: $C = C \cup C_{+}$, $I = I \cup I_{+}$, reflecting the update of the penetration tester's knowledge.
\end{definition}

\subsection{Penetration Testing Formulation}
The two central formalisms of this work include: a definition of the penetration testing problem and a definition of a strategy to guide the penetration testing actions.

\begin{definition}  
\label{def:pentest}
Given the definitions for the target system, the penetration test state, attacks and scans, a penetration test can be formally defined as a labelled transition system~\cite{GLABBEEK20013}, $T = \langle \PStates, \Attacks, \Sigma, \mathcal{T}, P_{0}, \PStates_{F} \rangle$ where:
\begin{itemize}
    \item { $\PStates$ is a set of penetration test states. A state $P \in \PStates$ is defined by definition~\ref{def:pentest.state},}
    \item { $\Attacks$ is a set of attacks. An attack $A \in \Attacks$ is defined by definition~\ref{def:attack}.}
    \item { $\Sigma$ is a set of scans. A scan $\sigma \in \Sigma$ is defined by definition~\ref{def:scan},}
    \item $\mathcal{T} \subseteq \PStates \times \Attacks \cup \Sigma  \times \PStates $ is a transition relation that models penetration test actions, we write:
    \begin{itemize}
        \item $P \xrightarrow{A} P^{\prime}$, when the selected action is an attack,  
        \item $P \xrightarrow{\sigma} P^{\prime}$, when the selected action is a scan,
    \end{itemize}
    \item $P_{0} \in \PStates$ is the initial penetration state,
    \item $\PStates_{F} \subseteq \PStates$ is a set of terminal states.
\end{itemize}

  We define a penetration test strategy as a tuple, $S = \langle S_{M}, S_{M_{0}}, S_{U}, S_{\mathcal{N}} \rangle$ where:
  \begin{itemize}
  \item $S_{M}$ is a finite set of memory elements,
    \item $S_{M_{0}}$ is the initial memory mapping, 
    \item $S_{U}: S_{M} \times \PStates \to M$ is a memory update function, where $S_{U}(S_{M},P)$ returns the updated contents of $S_{M}$ given the current state $P$,
    \item $S_{N}: S_{M} \times \PStates \to \mathcal{T} $ is a next move function, which returns the next $t \in \mathcal{T}$ to be taken based on the current penetration test state and memory contents.
  \end{itemize}
A strategy induces a (possibly infinite) transition path $\varpi = t_{s}, \dots, t_{n}$, starting in a state $P_{s} \in \mathcal{P}$ and ending in $P_{n} \in \mathcal{P}$ by repeatedly applying $S_{U}$ followed by $S_{N}$. Each such application produces the next transition $t \in \mathcal{T}$ to be taken in $T$. 
Intuitively, a penetration test strategy is successful if it leads to one of the terminal states in $\mathcal{P}_{F}$ within a finite number of steps, i.e., if starting from $P_{0} \in \mathcal{P}$ and $S_{M_{0}}$ and applying $S_{U}$ and $S_{N}$ a finite amount of times results in a state $P_{f} \in \mathcal{P}_{F}$. 

Penetration tests can have multiple goals. In our formulation we model a goal as a predicate that holds over terminating states in $\mathcal{P}_{F}$.
The predicate can range over any element of the penetration testing state, including the architectural properties of the components and interfaces.

The typical goal, however, is to discover as much of the system as possible and gain control of as much of the system as possible.
This goal is formalized as follows:
\begin{align} 
    \neg \exists A \mid  K \cup K_{A} \neq K \label{eq:attack} \\
    \neg \exists \sigma \mid  C \cup C_{+} \neq C 
    \label{eq:scan:components}\\ 
    \neg \exists \sigma \mid  I \cup I_{+} \neq I 
    \label{eq:scan:interfaces}
\end{align}
\end{definition}
In other words, equation~\ref{eq:attack} means that there is no attack left that will increase the penetration tester's capabilities if it succeeds. Similarly, equations~\ref{eq:scan:components} and ~\ref{eq:scan:interfaces} mean that there should be no scan left that will allow the penetration tester to learn about a new component  
or interface.

\subsection{RL and PDDL to our problem definition }

\subsubsection*{RL formalisms}
  The closest related work to ours that uses reinforcement-learning to achieve black-box penetration testing is~\cite{Zhou2019}.
  We informally show how to transform its formulation of the penetration testing problem to our formulation.
  The precise mapping is given in the appendix. 
  Other related work that utilizes reinforcement learning, e.g., ~\cite{9394285,9229752,app11198823}
  can be mapped following a similar process.
  In \cite{Zhou2019}, a penetration test is given by an Markov decision process~\cite{puterman2014markov}. 
  The state space is described by a number of machines. A machine is identified by an identifier, its operating system, a set of open ports, a set of known services and its exploitation state.
  The action space corresponds to a set of scans and a set of attacks. As is common for reinforcement-learning, a probability and a reward are associated with each action.
  The correspondence to our formulation is as follows: (i) each machine corresponds to a component with its operating system and exploitation state being component properties,
  (ii) each open port corresponds to an interface with known services being interface properties, (iii) scans and attacks make up our scans and attacks with the sole distinction
  that the model in~\cite{Zhou2019} only supports one step attacks and (iv) the action probabilities and rewards make up the memory elements of a strategy.

  \subsubsection*{PDDL formalisms}
    PDDL-based formulations are also easily mapped to our problem statement. As an example, we show how the PDDL-based formulation in ~\cite{DBLP:journals/corr/ObesSR13} can informally mapped to our definition.
    \cite{DBLP:journals/corr/ObesSR13} defines a classical AI-temporal planning problem~\cite{Fox2003PDDL21AE}.
    Its main elements are: (i) predicates that correspond to assets and their properties, (ii) actions that correspond to exploitation actions, i.e., attacks and (iii) a goal that corresponds to a set of satisfied predicates at
    the end of the plan. The mapping to our problem definition is straightforward based on the following observation: each predicate refers to either a host or a port, and hence can be mapped to a component or interface property in our formulation. Exploitation actions can be also directly mapped to attacks in our model, by mapping their preconditions to the set of required capabilities $K_{0}$ of each attack, and their effects to the set of gained capabilities $K_{A}$.
    The goal can be directly mapped to our goal after the predicates and actions have been
    translated.
    Note that \cite{DBLP:journals/corr/ObesSR13} is a white-box penetration testing approach, thus the sets of components $\mathcal{C}$ and interfaces $\mathcal{I}$ are considered to be known i.e., they are given by the problem definition. 
    The two solving strategies involve off-the shelf planners, namely metric-FF~\cite{Hoffmann} and  SGPlan~\cite{Chen2006TemporalPU}, both of which are principally forward search planners that employ heuristics. SGPlan in addition to heuristics employs a problem decomposition approach.

\section{\toolname}
\label{sec::architecture}
In this section we present an general self-adaptive system, i.e., the architecture and high level algorithms required to automate penetration testing as defined in Section~\ref{sec:problem}.
First we discuss each adaptation phase and its mapping to the elements of the previous section.
The following subsection describes the specific decisions made in our concrete implementation of \toolname that utilizes utility theory for its decision making.

The general architecture is shown in Fig.~\ref{Fig:General.Architecture}.
The managing subsystem acts as the decision-making engine that guides the penetration testing process while the managed subsystem is reconfigurable and provides the required tools to implement the decisions made by the managing system.
The managed system comprises a set of scanners, penetration testing and attack automation, and post-exploitation tools that are configured, coordinated and controlled by the managing subsystem which is implemented as an autonomic manager with the standard MAPE-K architecture. The tools that make up the managed system can be operated concurrently, i.e., multiple scans, attacks and post-exploitation operations may be active at any time. When a scan terminates the monitoring component gets notified by an event-based callback. Likewise, the execution component gets notified through a callback when an exploitation or post-exploitation action terminates. 

\begin{figure}
  \centering
  \includegraphics[scale=0.60]{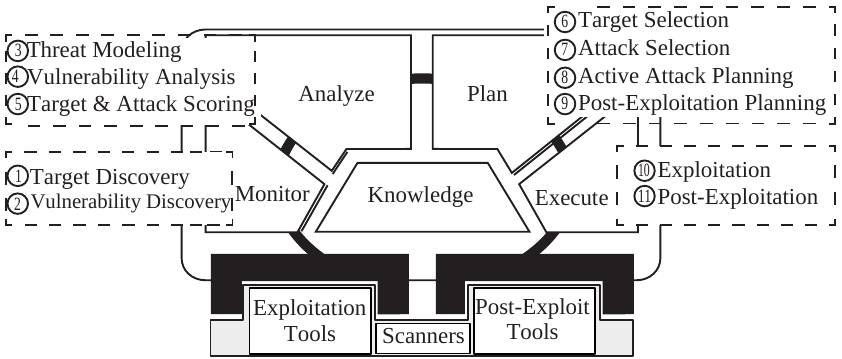}
  \caption[justification=centering]{Automated Penetration Testing Architecture}
  \label{Fig:General.Architecture}
\end{figure}

\subsubsection*{Knowledge-base}
The knowledge-base functions as the locus that provides the information required at runtime to implement and support the phases of the MAPE-K loop.
It stores information including: the penetration test state, the attack and scan repertoires and the strategy to guide the attack and scan selection.
Formally, the knowledge-base can be defined, using the elements presented in section~\ref{sec:problem}, as a tuple: $\mathcal{K} = \langle P, \Attacks, \Sigma, S  \rangle$.

The knowledge-base contains information that includes the penetration test state $P$, i.e., the part of the security-informed architecture, $\mathcal{S}$ that has been discovered.
The attack and scan repertoires amount to a set of attack tactics and a set scanning techniques.
In reference to definition~\ref{def:pentest}, the attack repertoire corresponds to $\Attacks$, and the scan repertoire corresponds to $\Sigma$.
An attack tactic is a sequence of attack techniques, each being an attack template, and which can be realized
by specifying the required concrete parameters. For example, to realize an attack technique that attempts to crack the weak password that resides in the WebServer component of the running example in Fig.~\ref{fig:running.example}, we need to specify: (i) a corresponding http form cracker tool, (ii) the IP and port of the WebServer's HTTP interface and (iii) the target form. Similarly to attack techniques, scanning techniques correspond to scan templates which need to be realized by providing a set of parameters. Attack techniques are realized by running exploitation tools while scan techniques are realized by running either scanners, e.g., network or vulnerability scanners or post-exploitation tools, e.g., credential or process discovery tools.
Table~\ref{tab:knowledgebase} shows an overview of the sets and subsets that the knowledge base maintains to keep track of the attack and scan execution history, and the exploitation state of each component. Note that $\Attacks_{\top}, \Attacks_{\bot}$ and $\Attacks_{*}$ are disjoint sets and $\Attacks \setminus \{ \Attacks_{\top} \cup \Attacks_{\bot} \cup \Attacks_{*} \}$ is the set of available attacks that can be selected to be tried next. Similarly, $\Sigma_{\top}$  and $\Sigma_{*}$ are disjoint and $\Sigma \setminus \{ \Sigma_{\top}  \cup \Sigma_{*} \} $ is the set of scans from which the next scan is to be selected from and $C \setminus \{ C_{\top}  \cup C_{*} \} $ is the set of components the next target component can be selected from.

\begin{table}[ht]
  \caption{Knowledge-base concepts and sets}
  \label{tab:knowledgebase}
  \centering
  \small
    \let\Tabular\tabular
    \def\tabular{\Tabular}
    \scalebox{0.8}{
    \begin{tabular}{cccccc}
      \toprule
      \textbf{Concept} &  \textbf{Set} & \textbf{Element} & \textbf{Active} & \textbf{Successful/Exploited} & \textbf{Failed} \\
      \midrule
      Component & $C$ & $c$ & $C_{*}$ & $C_{\top}$ & -   \\
      \midrule
      Attack & $\Attacks$ & $A$ & $\Attacks_{*}$ & $\Attacks_{\top}$ & $\Attacks_{\bot}$   \\
      \midrule
      Scan & $\Sigma$ & $\sigma$ & $\Sigma_{*}$ & $\Sigma_{\top}$ & -   \\
     \bottomrule
    \end{tabular}
    }
\end{table}

\subsubsection*{Monitoring}
The monitoring phase is responsible for updating the architecture's knowledge  in terms of targets (components that can be attacked) and possible attacks(either previously unknown vulnerabilities or previously unknown vulnerable interfaces). In particular, the monitoring phase utilizes two kinds of probes: (i) remote probes (ii) and local probes. Remote probes are deployed in the managed system and perform remote scanning through the network, while 
local probes are installed into the target components once they have been
compromised and perform both local and remote scanning. Local scanning amounts to scanning the component for interfaces that are not necessarily
exposed or accessible to the managed system, while remote scanning amounts to remote scanning through the compromised component to access networks that
the managed system is not part of.
In reference to definition~\ref{def:pentest}, the monitoring phase updates the penetration test state $P$, based on newly discovered components or interfaces of the target architecture $\mathcal{S}$.
The high-level execution of the monitoring phase is shown in \textbf{Adaptation Phase 1:} Monitoring. monitor\_run is responsible for starting a new monitoring phase, it receives a new set of
  scans to be performed, adds each scan to the set of active scans (line 3) and starts each scan in
  order (line 4). When a scan completes its operation, monitor\_control is invoked: first, the sets of active and completed scans are updated (lines 8 and 9), second, the penetration test state is updated with the scan results (lines 10 and 11) and finally a new analysis phase is invoked if either there exists a scan that has not been performed yet or one is running, or there exists an attack that has not been tried or one that is running. (lines 12-14).

\begin{algorithm}[htb]
\SetAlgorithmName{Adaptation Phase}{}{}
\DontPrintSemicolon 
\SetKwProg{Fn}{Function}{ }{end}
\small
\caption{Monitoring}
\label{adapt:monitoring}
\begingroup
\Fn{\textup{monitor\_run}($N_{\sigma}, \mathcal{K}$)}{
\KwIn{Set of scans to be started $N_{\sigma}$}
\KwIn{The knowledge-base $\mathcal{K} = \langle P, \Attacks, \Sigma, S  \rangle$}
  \For{ $\sigma \in N_{\sigma}$}{
    \tcp{\scriptsize{Add scan to active scans and start it}}
    $\Sigma_{*} = \Sigma_{*} \cup \{ \sigma \} $ \;
    \textup{scan\_start($\sigma$)} \;
  }
}
  \endgroup

\begingroup
\Fn{\textup{monitor\_control}($\sigma, \mathcal{K}$)}{
\KwIn{A completed scan $\sigma$}
\KwIn{The knowledge-base $\mathcal{K} = \langle P, \Attacks, \Sigma, S  \rangle$}
  \tcp{\scriptsize{Update active and completed scans}}
  $\Sigma_{*} = \Sigma_{*} \setminus \{ \sigma \} $ \;
  $\Sigma_{\top} = \Sigma_{\top} \cup \{ \sigma \} $ \;
  \tcp{\scriptsize{Update pentest state with scan results}}
  $C = C \cup C_{+}$\;
  $I = I \cup I_{+}$\;
  \tcp{\scriptsize{A scan is running or available or an attack is running or has not been tried}}
  \If{$ \Sigma_{*} \neq \varnothing \vee \exists \sigma : \sigma \notin \Sigma_{\top} \vee$ \\ \hspace{0.73em}  $\Attacks_{*} \neq \varnothing \vee \exists A : A \notin (\Attacks_{\top} \cup \Attacks_{\bot})$ }{
\textup{analysis\_run($\mathcal{K}$)}\;
}
}
\endgroup
\end{algorithm}

\subsubsection*{Analysis}
The analysis phase is responsible for attack modeling and vulnerability analysis.
An attack model is constructed and kept up to date based on the results of the monitoring phase.
The attack model is then utilized for vulnerability analysis which evaluates how vulnerable each component and interface are estimated to be and additionally performs attack prioritization to filter out the attacks that are not viable given the known state of the target system and rank the available attacks based on their likelihood of being effective. The results of the vulnerability analysis are a target and an attack ranking.
The attack model, and target and attack ranking make up the memory elements $S_{M}$ of $S$.
These rankings form the basis for selecting which scans(s) and attack(s) will be chosen to be performed next.
The high-level execution of the analysis phase is shown in \textbf{Adaptation Phase 2:} Analysis.
The analysis's main responsibilities are to retrieve and update the memory elements $S_{M}$ of the penetration test strategy $S$ and to invoke the memory update function $S_{U}$ (line 2) which in turn generates the information needed for the planning component to select the next scan(s) and attack(s) to be performed.

\begin{algorithm}[htb]
\SetAlgorithmName{Adaptation Phase}{}{}
\DontPrintSemicolon 
\SetKwProg{Fn}{Function}{ }{end}
\small
\caption{Analysis}
\label{adapt:analysis}
\begingroup
\Fn{\textup{analysis\_run}($\mathcal{K}$)}{
\KwIn{The knowledge-base $\mathcal{K} = \langle P, \Attacks, \Sigma, S  \rangle$}

  $S_{M} = S_{U}(S_{M},P)$\;
  \textup{planning\_run($\mathcal{K}$)} \;
  
}
\endgroup  
\end{algorithm}

\subsubsection*{Planning}
The planning phase has two main goals, to select the target components of the current adaptation cycle and to select the attacks and scans to be performed against them. The planning phase takes into account the target and attack ranking computed in the previous phase, the current estimated state of the target system and the availability of the managed system's resources and decides the next targets for attacks and scans, in addition to their parameters.
The planning phase's high-level operation is shown in \textbf{Adaptation Phase 3:} Planning.
In connection to the problem presented in Section.~\ref{sec:problem}, planning chooses the next attacks and scans to be performed, i.e., the penetration test actions in $\mathcal{T}$ by applying the next move action $S_{N}$(line 2).

\begin{algorithm}[htb]
\SetAlgorithmName{Adaptation Phase}{}{}
\DontPrintSemicolon 
\SetKwProg{Fn}{Function}{ }{end}
\small
\caption{Planning}
\label{adapt:planning}
\begingroup
\Fn{\textup{planning\_run}($\mathcal{K}$)}{
\KwIn{The knowledge-base $\mathcal{K} = \langle P, \Attacks, \Sigma, S  \rangle$}
  \tcp{\scriptsize{Get new attacks and scans according to $S$}}
  $N_{\Attacks}, N_{\sigma} = S_{N}(S_{M},P)$\;
  \textup{execution\_run($N_{\Attacks}, N_{\sigma}, \mathcal{K}$)} \;
}
\endgroup  
\end{algorithm}

\subsubsection*{Execution}
The execution phase configures and controls the tools that comprise the
managed system based on the results of the planning phase. The tools themselves and their configuration can be broadly categorized into two categories: (i) exploitation tools, which include: exploitation frameworks, crackers, injection discovery tools, etc and (ii) post-exploitation tools which include tools used for service and process discovery, lateral movement, persistence, command and control, exfiltration etc. Execution performs the attack(s) in $\Attacks$ or scan(s) in $\Sigma$ decided by the planning phase.
The high level operation of the execution phase is shown in \textbf{Adaptation Phase 4:} Execution. execution\_run  updates the active attack and scan sets (lines 3 and 7) and starts the
  corresponding attacks and scans (lines 4 and 8). execution\_control runs when an attack has concluded.
  It updates the set of active attacks (line 12) and the set of successful or failed attacks (lines 14 or 17). If the attack succeeded, then it additionally updates the penetration tester's capabilities (line 15). If the penetration testing goal has been met, the loop exits (lines 19 and 21, otherwise a new monitoring phase is started (line 22).

\begin{algorithm}[htb]
\setcounter{algocf}{3}
\SetAlgorithmName{Adaptation Phase}{}{}
\DontPrintSemicolon 
\SetKwProg{Fn}{Function}{ }{end}
\small
\caption{Execution}
\label{adapt:execution}
\begingroup
\Fn{\textup{execution\_run}($N_{\Attacks},N_{\sigma}, \mathcal{K} $)}{
\KwIn{The set of new attacks to be started $N_{\Attacks}$}
\KwIn{The set of new scans to be started $N_{\sigma}$}
\KwIn{The knowledge-base $\mathcal{K} = \langle P, \Attacks, \Sigma, S  \rangle$}
  \tcp{\scriptsize{Add attack to active attacks and start it}}
  \For{ $A \in N_{\Attacks}$}{
    $\Attacks_{*} = \Attacks_{*} \cup \{ A \} $ \;
    \textup{attack\_start($A$)} \;
  }
  \tcp{\scriptsize{Add scan to active scans and start it}}
  \For{ $\sigma \in N_{\Sigma}$}{
    $\Sigma_{*} = \Sigma_{*} \cup \{ \sigma \} $ \;
    \textup{scan\_start($\sigma$)} \;
  }
  
}
\endgroup

\begingroup
\Fn{\textup{execution\_control}($A, \mathcal{K}$)}{
\KwIn{A completed attack $A$}
\KwIn{The knowledge-base $\mathcal{K} = \langle P, \Attacks, \Sigma, S  \rangle$}
  \tcp{\scriptsize{Update active attacks  }}
  $\Attacks_{*} = \Attacks_{*} \setminus \{ A \} $ \;
  
  \uIf{\textup{successful($A$)}}{
    \tcp{\scriptsize{Update successful attacks }}
    $\Attacks_{\top} = \Attacks_{\top} \cup \{ A \} $ \;
    \tcp{\scriptsize{Update pentest state with new capabilities}}
    $K = K \cup K_{A}$\;
  }
  \Else{
    \tcp{\scriptsize{Update failed attacks}}
    $\Attacks_{\bot} = \Attacks_{\bot} \cup \{ A \} $ \;  
  }
  \tcp{\scriptsize{No attacks or scans active or left to try}}
  \uIf{$\Attacks \setminus (\Attacks_{\top} \cup \Attacks_{\bot} \cup \Attacks_{*}) = \varnothing$ $\wedge$  $ \Sigma \setminus (\Sigma_{\top} \cup \Sigma_{*}) = \varnothing$ }{
    \tcp{\scriptsize{Pentest complete}}
    \textup{\textup{write\_report\_and\_exit()}}\;
  } 
  \Else{
    \tcp{\scriptsize{Start a new adaptation loop}}
    \textup{\textup{monitor\_run($\varnothing,\mathcal{K}$)}}\;
  }
}
\endgroup

\end{algorithm}

\subsection{Instantiation Targeting Hosts and Services}
\label{subsec:concrete.arch}
To motivate the feasibility of our general design in a realistic setting, we present an instantiation of the generic architecture and the required (design-specific) algorithms.
This instantiation targets systems with hosts, services exposed by these hosts (or processes running on them), and interactions over a local area network.
To that end, a number of decisions are required
with respect to: (i) the tools that are used by the managed system or are deployed to the targets during post-exploitation, (ii) the attack modeling methodology, (iii) the vulnerability analysis approach and (iv) the target and scan and attack selection methodology.

\subsubsection*{Probes, Exploitation and Post-exploitation Tools}
We have selected a representative set of widely-used tools that are familiar to penetration testers in addition to basic system utilities that can be used as multi-purpose tools for multiple scanning and post-exploitation tasks.
Table~\ref{tab:tools} in the Appendix shows a partial-list of  the tools that make up our managed system which are used to craft our attacks, in addition to the probes and post-exploitation tooling that reside either in the managed system or are deployed on the exploited target hosts.

\subsubsection*{Attack Modeling}
Instead of an explicit attack model, e.g., an attack graph or an attack tree, we utilize the security-informed architecture of the target system as an implicit attack model. We maintain an architectural model of the target system's known components, their interfaces and their interactions, in addition to a number of
relevant security properties, i.e., the security-informed architecture of the target system, to keep track of the penetration test state and to aid in vulnerability analysis. 
The security-informed architecture is kept in the knowledge base and is continuously updated at runtime.

To describe the possible attacks, we develop an attack repertoire, which we also store in the knowledge base, from which attacks can be chosen, instantiated and attempted against a target host and service.
The attack repertoire comprises a number of attack tactics. 
An attack tactic chains together multiple attack techniques with a common goal. Each technique instantiation is an attack step. For example, an attack tactic with the goal to gain access to an administrative management interface of a web server might involve the following attack steps: (i) employ skipfish to enumerate the accessible URLs, followed by (ii) an NSE discovery script to identify the administrator access form and as a final step (iii) employ hydra to brute force the form and gain access.

Our architecture automates the coordination, runtime decision making, synchronization and tool operation aspects of penetration-testing based on the exploitation capabilities provided by the attack techniques that make up the tactics of the attack repertoire and the detection capabilities provided by the scans that make up the scan repertoire. In turn, the effectiveness of the attack techniques and scans depends on the available capabilities of the underlying exploitation, post-exploitation and scanning tools. Hence, the architecture will not be able to perform any attacks on hosts with services or processes that cannot be detected by the scans in the scan repertoire or that are not present in a step of an attack tactic. That is to say, we automate the selection and application of previously known, generic and widely applicable attacks rather than invent new attack techniques.

\subsubsection*{Vulnerability Analysis}
To perform vulnerability analysis, we employ utility-based decision making to determine how vulnerable each host is and which scans and attacks would be most effective against each host. The outcome of this process is the ranking of targets, scans and attacks based on their utility. 
Table~\ref{tab:target.properties} shows the properties stored in the attack model that are considered as factors to decide the target ranking, while Table~\ref{tab:attack.properties} shows the properties that are considered to decide the attack ranking and Table~\ref{tab:scan.properties} shows the properties considered to decide the scan ranking.
For brevity, we only show the factors and the corresponding
weights. The full details, including the value functions, are available in Tables ~\ref{tab:target.properties.full}, ~\ref{tab:attack.properties.full} and ~\ref{tab:scan.properties.full} in the Appendix.
The analysis phase, therefore, implements utility calculation and ranking following Equation~\ref{eq:utility}, provided that $S_{M} = \langle W_{C}, W_{\Attacks}, W_{\Sigma}, V_{C}, V_{\Attacks}, V_{\Sigma}, U_{C}, U_{\Attacks}, U_{\Sigma} \rangle$
where:
\begin{itemize}
    \item $W_{C}$, $W_{\Attacks}$ and $W_{\Sigma}$ correspond to vectors that encapsulate the
    weight entries in Table~\ref{tab:target.properties}, Table~\ref{tab:attack.properties}, and Table~\ref{tab:scan.properties},
    \item $V_{C}$, $V_{\Attacks}$ and $V_{\Sigma}$ correspond to vectors that encapsulate the value entries in the Table~\ref{tab:target.properties.full}, Table~\ref{tab:attack.properties.full}, and Table~\ref{tab:scan.properties.full},
    \item $U_{C}$, $U_{\Attacks}$ and $U_{\Sigma}$ are utility vectors, one for each target, attack and scan that make up the rankings returned by the analysis.
\end{itemize}
For every available target, attack and scan (lines  2, 6 and 10), we calculate its utility according to equation~\ref{eq:utility} (lines 3, 7 and 11).
Once the utility of every option, i.e., attack, scan or target has been computed, we rank them by sorting the corresponding array (lines 5, 9 and 13) and filtering out entries with zero utility.

\renewcommand{\thealgocf}{}
\begin{algorithm}[htb]
\SetAlgorithmName{Memory Update}{0}{0}
\DontPrintSemicolon 
\SetKwProg{Fn}{Function}{ }{end}
\small
\caption{ ADAPT}
\label{adapt:strategyupdate}
\begingroup
\Fn{$S_{U}$($S_{M},P$)}{
\KwIn{The memory elements, $S_{M} = \langle W_{C}, W_{\Attacks}, W_{\Sigma}, V_{C}, V_{\Attacks}, V_{\Sigma}, U_{C}, U_{\Attacks}, U_{\Sigma} \rangle$ }
\KwIn{The penetration test state $P$}
  \For{ $c  \in C \setminus (C_{*} \cup C_{\top})$}{
   $U_{C}(c) = \underset{{w \in W_{C}, u \in U_{C}}}{\sum} w \cdot u(c)  $\;
  }
sort($U_{C}$) \;
\For{$A  \in \Attacks \setminus (\Attacks_{*} \cup \Attacks_{\top} \cup \Attacks_{\bot})$}{
$U_{\mathcal{A}}(A) = \underset{{w \in W_{\Attacks}, u \in V_{\Attacks}}}{\sum} w \cdot u(A)  $ \;
}
sort($U_{\mathcal{A}}$) \;
\For{$\sigma  \in \Sigma \setminus (\Sigma_{*} \cup \Sigma_{\top})$}{
$U_{\Sigma}(\sigma) = \underset{{w \in W_{\Sigma}, u \in V_{\Sigma}}}{\sum} w \cdot u(\sigma)  $ \;
}
sort($U_{\Sigma}$) \;
}
\endgroup  
\end{algorithm}

\begin{table}[ht]
   \caption{Target Decision Factors}
  \label{tab:target.properties}
  \footnotesize
  \centering
    \let\Tabular\tabular
    \def\tabular{\Tabular}
   \begin{tabular}{ccc}
      \toprule
    \textbf{Factor} & \textbf{Weight}  \\
      \midrule
       Number of Services(S) & 0.2  \\
       Number of Vulnerabilities(V) & 0.2  \\
       Number of Connections(C) & 0.2 \\
      Exploitation State  (E) & 0.4\\
      \bottomrule
    \end{tabular}
\end{table}

\begin{table}[ht]
\caption{Attack Decision Factors}
\label{tab:attack.properties}
\footnotesize
\centering
\begin{tabular}{cc}
\toprule
\textbf{Factor} & \textbf{Weight}   \\
\midrule
Attack Vector (V) & 0.3 \\

Attack Complexity (C) & 0.2  \\

Required Privileges (P) & 0.2 \\

User Interaction (U)  & 0.1  \\

Running Time  (T) & 0.2 \\
\bottomrule
\end{tabular}
\end{table}

\begin{table}[ht]
\caption{Scan Decision Factors}
\label{tab:scan.properties}
\footnotesize
\centering
\begin{tabular}{cc}
\toprule
\textbf{Factor} & \textbf{Weight}  \\
Scan Range(R) & 0.25 \\
Number of Targets (T) & 0.25 \\
Scan Duration(D) & 0.25 \\
Scan Complexity(C) & 0.25\\
\bottomrule
\end{tabular}
\end{table}

\subsubsection*{Planning}
The planning process is guided by the results of target and attack ranking, taking into account two additional factors: (i) a threshold for each of the number of concurrent targets ($C_{C}$) and attacks ($C_{A})$ and 
(ii) the attacks and post-exploitation tools already active.
We employ a basic resource allocation algorithm as shown in \textbf{Next Move(s):} ADAPT to select the next actions to be taken in the next execution phase, i.e., invoking $S_{N}$. The high-level description of the target allocation algorithm, target\_selection, is as follows: as long as the number of allowed concurrent targets ($C_{C})$ has not been exceeded (line 10), we traverse the ranking of the targets and add the next most valuable target (line 14), provided that: (i) it has not been fully exploited and (ii) it is not already in the list of active targets (line 13). Similarly, the attack allocation algorithm attack\_selection, will allocate an attack (line 28), if the number of allowed concurrent attacks $(C_{A}$) has not been exceeded (line 25), provided that: (i) it attacks an active target, (ii) it has not been attempted before, (iii) it is not in the list of active attacks for the target, and (iv) it does not  interfere with other running attacks (line 27).
In the interest of brevity, we do not give the scan\_selection function, as it is almost identical to attack\_selection 

\begin{algorithm}
\DontPrintSemicolon 
\SetAlgorithmName{Next Move(s)}{}{}
\DontPrintSemicolon 
\SetKwProg{Fn}{Function}{ }{end}
\small
\caption{ADAPT}
\label{adapt:nextmove}
\begingroup
\Fn{$S_{N}$($S_{M},P$)}{
\KwIn{The memory elements, $S_{M} = \langle W_{C}, W_{\Attacks}, W_{\Sigma}, V_{C}, V_{\Attacks}, V_{\Sigma}, U_{C}, U_{\Attacks}, U_{\Sigma} \rangle$ }
\KwIn{The penetration test state $P$}

$N_{C}$ = target\_selection($U_{C}$)\;
$N_{A}$ = attack\_selection($U_{\mathcal{A}}$)\;
$N_{\sigma}$ = scan\_selection($U_{\Sigma}$)\;
return $N_{A}$, $N_{\sigma}$ \;
}
\endgroup
\begingroup
\Fn{\textup{target\_selection($U_{C}$)}}{
\KwIn{target utility ranking, $U_{C}$}
$N_{C} = \varnothing$ \;
$C_{C}$ = $C_{C}$ - $|C_{*}|$ \;
$idx$ = 0 \;

\While{ $C_{C} > 0 \wedge idx < |U_{C}|$}{
$c$ = \textup{target\_from\_idx($U_{C}$,$idx$)} \;
\If{$c \notin (C_{\top} \cup C_{*})$} {
$N_{C}$ = $N_{C}$ $\cup$  \{$c$\} \;
$C_{C}$ = $C_{C}$ - 1 \;
}
$idx = idx + 1$\;
}
\Return $N_{C}$ \;
}
\endgroup

\begingroup
\Fn{\textup{attack\_selection}($U_{\mathcal{\Attacks}}$)}{
\KwIn{target utility ranking, $U_{\Attacks}$}
$N_{A} = \varnothing$ \;
$C_{A}$ = $C_{A}$ - $|\Attacks_{*}|$ \;
$idx$ = 0 \;
\While{ $C_{A} > 0 \wedge idx < |U_{\Attacks}|$}{
$A$ = \textup{attack\_from\_idx($U_{\Attacks}$,$idx$)}\;
\If{$A \notin (\Attacks_{\top} \cup \Attacks_{\bot} \cup \Attacks_{*}) \wedge \neg \textup{interferes}(A,\Attacks_{*}) \wedge \textup{target\_of}(A) \in C_{*}$}{
$N_{A} = N_{A} \cup \{A\}$ \;
$C_{A}$ = $C_{A}$ - 1 \;
}
$idx = idx + 1$\;
}
\Return $N_{A}$ \;
}
\endgroup  
\end{algorithm}


\begin{example}
To illustrate how vulnerability analysis and target and attack allocation work, we demonstrate the first adaptation cycle decisions for the system shown in Fig.~\ref{fig:running.example}. We consider the initial penetration testing state to be the state after a simple network scan has run where only the two internet facing components and the two internet facing interfaces have been identified. In other words, $C = \{WebServer, APIGate\}$ and $I = \{ HTTP, gRPC\}$. Both components have the same known number of services, vulnerabilities 
and connections and are in the same exploitation state, hence the utility of targeting either is: 0.22. 
Moreover, let us assume that the attack repertoire only 
contains two possible attack tactics: (i) HTTP Form Cracking targeting the WebServer and (ii) a custom exploit that takes advantage of the out of bounds write vulnerability in the deployed gRPC implementation on APIGate. Both attacks can be performed over the network, and do not require any privileges or user interaction. However, HTTP Form Cracking has a low attack complexity and long running time, while the custom exploit has a high attack complexity and quick running time. Thus the utility of employing  HTTP Form Cracking is: 0.863 and the utility of using the custom exploit is: 0.903. Applying the target and attack allocation algorithm for $C_{C},C_{A}$ = 1, will result in randomly picking between one of the two targets, since their utilities are equal. Attack allocation filters out attacks that are not against active targets,  hence the attack selected will be HTTP Form Cracking if the selected target is  WebServer and the custom exploit if the selected target is APIGate. For $C_{C}$, $C_{A}$ = 2 or higher, both targets and attacks would be selected.
\end{example}




\newcommand{\hochkomma}{$^{,}$}

\section{Evaluation}
\label{sec::evaluation}
We have implemented the architecture described in Section.~\ref{subsec:concrete.arch}, which we evaluate in this section.
We have designed the managed system with a plugin-based architecture, so that different tool plugins can provide the capabilities to support multiple attack techniques.
The tool source code is released under an open license\footnote{https://gitfront.io/r/anonymous-submitter/P2LRhxvh9L7z/ADAPT/}.
Due to ethical considerations, we do not release \emph{intrusive} tool plugins publicly, but we are committed to open science and will facilitate experiment replication and comparisons.

We evaluate our tool against Metasploitable2, Metasploitable3, and a VM-based network.
Metasploitable2 and Metasploitable3 are vulnerable VMs often used as first targets to test penetration testing tools.
The VM-based network has been developed at our university\footnote{University name to be provided in camera ready version} to serve as the lab part of a course in ethical hacking.
The goal of the experiments is to demonstrate that ADAPT can perform a penetration test against a realistic system autonomously, making all the relevant decisions to compromise the system within a reasonable number actions, with minimal overhead, and without access to any specific information about the system architecture or functionality beyond the information used to build the scan and attack repertoires. 

\subsection{Experiment Setup}
Metasploitable2 is a intentionally vulnerable VM running Ubuntu 8.04.
It exposes multiple vulnerable services including various non-password protected services(smtp, rlogin, rsh, rexec, nfs), various backdoors(vsftpd, unrealIRCd),
services with weak passwords(ssh, telnet, postgres, vnc), and services with well known vulnerabilities(httpd, javarmi, samba). Moreover, since the kernel version is outdated,
there are multiple privilege-escalation exploits available once local user access has been obtained.
Similarly, Metaspoitable3 has been designed and is available for the same purposes albeit featuring a higher degree of difficulty.
It comes in two flavors, as a Windows windows2008 or as an Ubuntu 14.04 server. we have opted to use the Windows version in our evaluation.
A multitude of vulnerable services, or services with weak passwords are also exposed by Metasploitable3, including for instance, struts, glassfish, mysql, iis-ftp, iis-http, etc.

The VM network setup comprises both Windows server 2k19 and Ubuntu 18.04 virtual machines, connected in a virtual network.
The VM network setup is used in an ethical hacking course at master's level for computer science students.
The course corresponds to 4 weeks of full-time studies and the lab part of it corresponds to 3 weeks full-time studies split into 16 calendar weeks.
Eleven flags are embedded in the system hosts. Gaining the flags, requires the usage of multiple different exploitation tools,
in addition to network and process discovery, privilege escalation, network sniffing, and lateral movement through the network.
To give a rough sense of the difficulty of the labs, which can serve as an estimate of the effort required to
perform manual penetration testing against our lab setup, we provide statistics of the time taken to capture the different flags
during spring 2024. The students were split in 39 groups of two. Fig.~\ref{Fig:Flag.Completion} shows the flag capture statistics per week.

\begin{figure}
\noindent \begin{tikzpicture}[scale=0.48]
  \foreach \y [count=\n] in {
  {     9 ,     29 ,      36 ,      36 ,     36 ,     36 ,      36 ,      36 ,     36 ,     36 ,      36 ,      36 ,     36 ,    36 ,     36 ,      36 },
{     7 ,     16 ,      31 ,      36 ,     36 ,     36 ,      36 ,      36 ,     36 ,     36 ,      36 ,      36 ,     36 ,    36 ,     36 ,      36 },
{    4 ,     17 ,      25 ,      31 ,     35 ,     38 ,      38 ,      38 ,     38 ,     38 ,      38 ,      38 ,     38 ,    38 ,     38 ,      38  },
{     3 ,      9 ,      15 ,      25 ,     34 ,     34 ,      34 ,      34 ,     34 ,     35 ,      35 ,      36 ,     36 ,    36 ,     36 ,      36 },
{     1 ,      7 ,      11 ,      19 ,     25 ,     29 ,      30 ,      31 ,     31 ,     31 ,      32 ,      33 ,     33 ,    33 ,     33 ,      33 },
{     0 ,      0 ,       0 ,       0 ,      0 ,      0 ,       1 ,       7 ,     12 ,     29 ,      30 ,      30 ,     31 ,    31 ,     34 ,      34 },
{     0 ,      0 ,       0 ,       0 ,      0 ,      0 ,       0 ,       0 ,      0 ,     12 ,      23 ,      28 ,     28 ,    29 ,     30 ,      31 },
{     0 ,      0 ,       0 ,       0 ,      0 ,      0 ,       0 ,       1 ,      1 ,     12 ,      19 ,      27 ,     27 ,    28 ,     30 ,      30 },
{     0 ,      0 ,       0 ,       0 ,      0 ,      0 ,       0 ,       0 ,      0 ,      2 ,       9 ,      20 ,     26 ,    27 ,     29 ,      31 },
{     0 ,      0 ,       0 ,       0 ,      0 ,      0 ,       0 ,       0 ,      0 ,      1 ,      10 ,      20 ,     26 ,    26 ,     27 ,      30 },
{    0 ,      0 ,       0 ,       0 ,      0 ,      0 ,       0 ,       0 ,      0 ,      0 ,       0 ,       3 ,      4 ,     4 ,      7 ,      18 },
    {},
    {},
    {},
    {},
    {},
  } {
    \ifnum \n<17
    \node[minimum size=4mm] at (\n, 0) {$\n$};
    \fi
    \foreach \x [count=\m] in \y {
      \node[fill=green!\col!white, minimum size=6mm, text=black] at (\m,-\n) {\x};}}

  \foreach \a [count=\i] in {$F_{1}$,$F_{2}$,$F_{3}$,$F_{4}$,$F_{5}$,$F_{6}$,$F_{7}$,$F_{8}$,$F_{9}$,$F_{10}$,$F_{11}$} {
    \node[minimum size=6mm] at (0,-\i) {\a};
  }
\end{tikzpicture}
\caption[justification=centering]{Groups that have captured each flag per week}
\label{Fig:Flag.Completion}
\end{figure}

\subsection{Initial Tool Setup}
We assume a black-box penetration testing set up; only
access to a network range where the hosts reside and the flag identifiers are provided to the tool.
Any host in the range can be targeted by scanners, offensive or penetration testing tools and once privileged access is gained can be modified to serve the attacker's needs, e.g.,
to launch further attacks against other valid targets in the range.
No information is provided regarding the possible services  on any of the hosts or any potential vulnerabilities.

\toolname implements the architecture presented in Section.~\ref{subsec:concrete.arch}, hence it uses the functions described in \textbf{Next Move(s)} ADAPT for target and attack selection. 
\toolname is equipped with a scan repertoire that is able to perform 8 scans and an attack repertoire with 41 attack tactics and 57 attack techniques. 
The scanning and exploitation capabilities provided by the repertoires directly affect the extent to which the penetration test succeeds. 
Thus, to showcase ADAPT's functionality,
we opted to use repertoires that contain the required capabilities
to successfully complete the individual exploits against the hosts of the VM network.
We do not measure ADAPT's success in terms of having a sufficient scan and attack repertoire to discover and exploit every vulnerability in each host independently,
but in terms of its ability to (i) successfully navigate the VM network, (ii) make the required runtime decisions to advance the penetration test state and (iii) operate the scanners, exploitation and post-exploitation tools with the appropriate context-dependent parametrization.
Setting up the attack and scan repertoires involves designing the attack tactics and selecting scans to be performed.
Designing an attack tactic involves selecting the attack techniques to be used, their sequence of execution and the corresponding utility values for each factor.

\subsection{Results}

\subsubsection*{Penetration Test Overview}
To demonstrate ADAPT's execution we show exploitation graphs, with similar semantics to logical attack graphs~\cite{MULVAL}. They have the following types of nodes: (i) ellipse nodes represent scans(prefixed with 'S'), exploitation (prefixed by 'E') and post-exploitation (prefixed with 'P') techniques, (ii) rectangle nodes represent interfaces(services or processes) and (iii) diamond nodes show gained capabilities.

\noindent \textbf{Metasploitable2 and Metasploitable3}
To perform a penetration test against  Metasploitable2, \toolname only needs to make use of a few tools which include: (i) nmap for service and vulnerability scanning and to bruteforce ftp credentials and http forms,  (ii) metasploit for exploitation and password cracking.
Fig.~\ref{Fig:Metasploitable2.Graph} shows the exploitation graph that is generated by merging the results of \emph{multiple executions} of \toolname.
A description of each node in the exploitation graph  and the mean running time for scans, exploitation and post-exploitation operations
can be found in table~\ref{tab:metasploitable2.detail} in appendix~\ref{Exploitation.Graph.Descriptions}.
Getting root permissions (shown as $C_{6}$ in Fig.~\ref{Fig:Metasploitable2.Graph}) on Metasploitable2 typically requires exploiting a single vulnerability and
can be achieved through 11 different interfaces, $I_{5}$ to $I_{15}$. Moreover, a regular user, msfadmin (shown as $C_{4}$), has passwordless sudo privileges, leading to a two step compromise.
The tool has additionally discovered a longer exploitation path by getting access through the vulnerable httpd service ($I_{4}$).
A CGI Argument vulnerability allows terminal access for user www-data ($C_{5}$).
The tool then runs a series of privilege escalation scanners, $S_{1}$, $S_{2}$ and $S_3$, each of which in turn discover a series of privilege escalation vulnerabilities shown as $C_{1}$, $C_{2}$ and $C_{3}$,
each of which can be exploited to escalate privileges.

The results of running our tool against Metasploitable3 can be seen in Fig.~\ref{Fig:Metasploitable3.Graph} and in table ~\ref{tab:metasplotable3.detail} in appendix~\ref{Exploitation.Graph.Descriptions}.
Similarly to Metasploitable2, the exploitation graph shows \emph{multiple runs} of our tool utilizing nmap and metasploit.
Metasploitable3 presents a harder challenge to fully compromise in comparison to its predecessor, typically, at least two attack techniques need to be employed to gain root privileges.
The most common exploitation scenario is having to crack the password of a service allowing access as a low-privileged user and performing an exploit that allows the tool to gain System or Administrator access.
Typically, that is achieved by exploiting the Eternal Blue vulnerability $E_{30}$ identified by running a post-exploitation scan $S_{1}$.
However, two services, Apache Struts and ManageEngine allow direct high privileged access once compromised.
An interesting exploitation path that does not lead to gaining administrator level privileges but does have security repercussions identified by our tool is exploiting the iis-http webserver to dump process memory.
This is a two step process, first a password cracker ($E_{17}$) is used to gain access to the web server's manager and then CVE2015-1635 is exploited through metasploit.

\begin{figure}[ht]
\begin{subfigure}[b]{0.45\textwidth}
\rotatebox{0}{
\scalebox{0.5}{
\begin{tikzpicture}[scale=1,every text node part/.style={align=center,draw,font=\Large},  every edge/.style={draw=black, very thick}, node distance=4cm,auto, initial text = {}]

  \node (S0) [circle,  minimum width = 0.5cm,  fill=gray!10, draw] at (0,3)  { {$S_{0}$}};
  \node (I0) [rectangle, minimum width = 0.5cm,  fill=gray!10, draw] at (-7.5,1)  { {$I_{0}$}};
  \node (I1) [rectangle, minimum width = 0.5cm,  fill=gray!10, draw] at (-6.5,1)  { {$I_{1}$}};
  \node (I2) [rectangle, minimum width = 0.5cm,  fill=gray!10, draw] at (-5.5,1)  { {$I_{2}$}};
  \node (I3) [rectangle, minimum width = 0.5cm,  fill=gray!10, draw] at (-4.5,1)  { {$I_{3}$}};
  \node (I4) [rectangle, minimum width = 0.5cm,  fill=gray!10, draw] at (-3.5,1)  { {$I_{4}$}};
  \node (I5) [rectangle, minimum width = 0.5cm,  fill=gray!10, draw] at (-2.5,1)  { {$I_{5}$}};
  \node (I6) [rectangle, minimum width = 0.5cm,  fill=gray!10, draw] at (-1.5,1)  { {$I_{6}$}};
  \node (I7) [rectangle, minimum width = 0.5cm,  fill=gray!10, draw] at (-0.5,1)  { {$I_{7}$}};
  \node (I8) [rectangle, minimum width = 0.5cm,  fill=gray!10, draw] at (0.5,1)  { {$I_{8}$}};
  \node (I9) [rectangle, minimum width = 0.5cm,  fill=gray!10, draw] at (1.5,1)  { {$I_{9}$}};
  \node (I10) [rectangle, minimum width = 0.5cm,  fill=gray!10, draw] at (2.5,1)  { {$I_{10}$}};
  \node (I11) [rectangle, minimum width = 0.5cm,  fill=gray!10, draw] at (3.5,1)  { {$I_{11}$}};
  \node (I12) [rectangle, minimum width = 0.5cm,  fill=gray!10, draw] at (4.5,1)  { {$I_{12}$}};
  \node (I13) [rectangle, minimum width = 0.5cm,  fill=gray!10, draw] at (5.5,1)  { {$I_{13}$}};
  \node (I14) [rectangle, minimum width = 0.5cm,  fill=gray!10, draw] at (6.5,1)  { {$I_{14}$}};
  \node (I15) [rectangle, minimum width = 0.5cm,  fill=gray!10, draw] at (7.5,1)  { {$I_{15}$}};

  \node (E0) [circle, inner sep=0.05cm,  fill=gray!10, draw] at (-7.5,0)  { {$E_{0}$}};
  \node (E1) [circle, inner sep=0.05cm,  fill=gray!10, draw] at (-6.5,0)  { {$E_{1}$}};
  \node (E2) [circle, inner sep=0.05cm,minimum width = 0.5cm,  fill=gray!10, draw] at (-5.5,0)  { {$E_{2}$}};
  \node (E3) [circle, inner sep=0.05cm, minimum width = 0.5cm,  fill=gray!10, draw] at (-4.5,0)  { {$E_{3}$}};
  \node (E4) [circle, inner sep=0.05cm, minimum width = 0.5cm,  fill=gray!10, draw] at (-3.5,0)  { {$E_{4}$}};
  \node (E5) [circle, inner sep=0.05cm, minimum width = 0.5cm,  fill=gray!10, draw] at (-2.5,0)  { {$E_{5}$}};
  \node (E6) [circle, inner sep=0.05cm, minimum width = 0.5cm,  fill=gray!10, draw] at (-1.5,0)  { {$E_{6}$}};
  \node (E7) [circle, inner sep=0.05cm, minimum width = 0.5cm,  fill=gray!10, draw] at (-0.5,0)  { {$E_{7}$}};
  \node (E8) [circle, inner sep=0.05cm, minimum width = 0.5cm,  fill=gray!10, draw] at (0.5,0)  { {$E_{8}$}};
  \node (E9) [circle, inner sep=0.05cm, minimum width = 0.5cm,  fill=gray!10, draw] at (1.5,0)  { {$E_{9}$}};
  \node (E10) [circle, inner sep=0.01cm, minimum width = 0.5cm, minimum height = 0.7cm,  fill=gray!10, draw] at (2.5,0)  { {$E_{10}$}};
  \node (E11) [circle, inner sep=0.01cm, minimum width = 0.5cm, minimum height = 0.7cm,   fill=gray!10, draw] at (3.5,0)  { {$E_{11}$}};
  \node (E12) [circle, inner sep=0.01cm, minimum width = 0.5cm, minimum height = 0.7cm, fill=gray!10, draw] at (4.5,0)  { {$E_{12}$}};
  \node (E13) [circle, inner sep=0.01cm, minimum width = 0.5cm, minimum height = 0.7cm,  fill=gray!10, draw] at (5.5,0)  { {$E_{13}$}};
  \node (E14) [circle, inner sep=0.01cm, minimum width = 0.5cm, minimum height = 0.7cm,  fill=gray!10, draw] at (6.5,0)  { {$E_{14}$}};
  \node (E15) [circle, inner sep=0.01cm, minimum width = 0.5cm, minimum height = 0.7cm,  fill=gray!10, draw] at (7.5,0)  { {$E_{15}$}};

  \node (C4) [diamond, inner sep=0.005cm, minimum width = 0.5cm,  fill=gray!10, draw] at (-7,-1)  { {$C_{4}$}};
  \node (C5) [diamond, inner sep=0.005cm, minimum width = 0.5cm,  fill=gray!10, draw] at (-3.5,-1.25)  { {$C_{5}$}};
  \node (E16) [circle, inner sep=0.05cm, minimum width = 0.5cm,  fill=gray!10, draw] at (-7,-3.25)  { {$P_{0}$}};

  \node (S1) [circle, inner sep=0.05cm, minimum width = 0.5cm,  fill=gray!10, draw] at (-4.75,-2.5)  { {$S_{1}$}};
  \node (S2) [circle, inner sep=0.05cm, minimum width = 0.5cm,  fill=gray!10, draw] at (-3.5,-2.5)  { {$S_{2}$}};
  \node (S3) [circle, inner sep=0.05cm, minimum width = 0.5cm,  fill=gray!10, draw] at (-2.25,-2.5)  { {$S_{3}$}};
  \node (C1) [diamond, inner sep=0.005cm, minimum width = 0.5cm,  fill=gray!10, draw] at (-4.75,-3.75)  { {$C_{1}$}};
  \node (C2) [diamond, inner sep=0.005cm, minimum width = 0.5cm,  fill=gray!10, draw] at (-3.5,-3.75)  { {$C_{2}$}};
  \node (C3) [diamond,inner sep=0.005cm, minimum width = 0.5cm,  fill=gray!10, draw] at (-2.25,-3.75)  { {$C_{3}$}};
  \node (E17) [circle, inner sep=0.05cm, minimum width = 0.5cm,  fill=gray!10, draw] at (-4.75,-5)  { {$E_{16}$}};
  \node (E18) [circle, inner sep=0.05cm, minimum width = 0.5cm,  fill=gray!10, draw] at (-3.5,-5)  { {$E_{17}$}};
  \node (E19) [circle, inner sep=0.05cm, minimum width = 0.5cm,  fill=gray!10, draw] at (-2.25,-5)  { {$E_{18}$}};

  \node (C6) [diamond, inner sep=0.005cm, minimum width = 0.5cm,  fill=gray!10, draw] at (2.25,-7)  { {$C_{6}$}};

  \path[->] 
  (S0) edge[bend right=20] node {} (I0)  
  (S0) edge[bend right=19] node {} (I1)
  (S0) edge[bend right=18] node {} (I2)
  (S0) edge[bend right=17] node {} (I3)
  (S0) edge[bend right=16] node {} (I4)
  (S0) edge[bend right=15] node {} (I5)
  (S0) edge node {} (I6)
  (S0) edge node {} (I7)
  (S0) edge node {} (I8)
  (S0) edge node {} (I9)
  (S0) edge[bend left=15] node {} (I10)
  (S0) edge[bend left=16] node {} (I11)
  (S0) edge[bend left=17] node {} (I12)
  (S0) edge[bend left=18] node {} (I13)
  (S0) edge[bend left=19] node {} (I14)
  (S0) edge[bend left=20] node {} (I15)
  (I0) edge node {} (E0)
  (I1) edge node {} (E1)
  (I2) edge node {} (E2)
  (I3) edge node {} (E3)
  (I4) edge node {} (E4)
  (I5) edge node {} (E5)
  (I6) edge node {} (E6)
  (I7) edge node {} (E7)
  (I8) edge node {} (E8)
  (I9) edge node {} (E9)
  (I10) edge node {} (E10)
  (I11) edge node {} (E11)
  (I12) edge node {} (E12)
  (I13) edge node {} (E13)
  (I14) edge node {} (E14)
  (I15) edge node {} (E15)
  (E0) edge node {} (C4)
  (E1) edge node {} (C4)
  (E4) edge node {} (C5)
  (C4) edge node {} (E16)
  (C5) edge node {} (S1)
  (C5) edge node {} (S2)
  (C5) edge node {} (S3)
  (S1) edge node {} (C1)
  (S2) edge node {} (C2)
  (S3) edge node {} (C3)
  (C1) edge node {} (E17)
  (C2) edge node {} (E18)
  (C3) edge node {} (E19)
  (E17) edge[bend right=20] node {} (C6)
  (E18) edge[bend right=15] node {} (C6)
  (E19) edge node {} (C6)
  (E5) edge[bend right=10] node {} (C6)
  (E6) edge[bend right=8] node {} (C6)
  (E7) edge node {} (C6)
  (E8) edge node {} (C6)
  (E9) edge node {} (C6)
  (E10) edge node {} (C6)
  (E11) edge node {} (C6)
  (E12) edge node {} (C6)
  (E13) edge node {} (C6)
  (E14) edge[bend left=8] node {} (C6)
  (E15) edge[bend left=10] node {} (C6)
  (E16) edge[bend right=35] node {} (C6)
  ;

  
\end{tikzpicture}
}
}
\end{subfigure}
\caption[justification=centering]{The exploitation graph for Metasploitable2}
\label{Fig:Metasploitable2.Graph}
\end{figure}

\begin{figure}[ht]
\input{figures/metasploitable3}
\caption[justification=centering]{The exploitation graph for Metasploitable3}
\label{Fig:Metasploitable3.Graph}
\end{figure}

\noindent \textbf{VM Network}
The exploitation graph for a \emph{single run} of our tool against the VM network setup is shown in Fig.~\ref{Fig:Lab.Environment.Graph}.
The lab involves multiple network hosts, we use the following notation for interfaces: "$H_{n}:I_{m}$",
where $H$ and $I$ signify host and interface respectively and $n$ and $m$ are natural numbers.
We show the flags found as capabilities, identified by an initial 'F' character. Since the flags in the VM network are actively being used as examination material in multiple courses given by higher education institutions, we cannot present or describe the exploitation steps in detail.
Nevertheless, we have chosen to add its penetration testing results to demonstrate our tool's capabilities in tasks not covered by Metasploitable2 and Metasploitable3,
namely: lateral movement, prioritizing among different hosts and exploitation in depth. Contrary to Metasploitable2 and Metasploitable3, where there is only a single host
and thus lateral movement and target prioritization are impossible, there are multiple hosts in the lab setup and lateral movement is a requirement to get all flags.
Moreover, the longest attack path for Metasploitable2 requires 4 attacker actions($S_{0},E_{4},S_{2},E_{17}$), involving the exploitation of 2 interfaces.  At most 7 attacker actions ($S_{0}, E_{2}, P_{0}, E_{19}, E_{28}, S_{1}, E_{30}$), are required to compromise Metasploitable3, involving the exploitation of 2 interfaces.
The longest attack path depth in the VM network is 12 actions($S_{0}, E_{0}, E_{1}, E_{9}, E_{10}, E_{11}, E_{12}, E_{13}, \allowbreak S_{2},\allowbreak  E_{15}, E_{16}, E_{17}$ ) requiring the exploitation of 4 interfaces($H_{0}:I_{0}, H_{0}:I_{1}, H_{4}:I_{0}, H_{4}:I_{1}$) in two different hosts. 
As seen in the exploitation graph, our tool is capable of finding all 11 flags; it is additionally able to get root privileges on all network hosts.
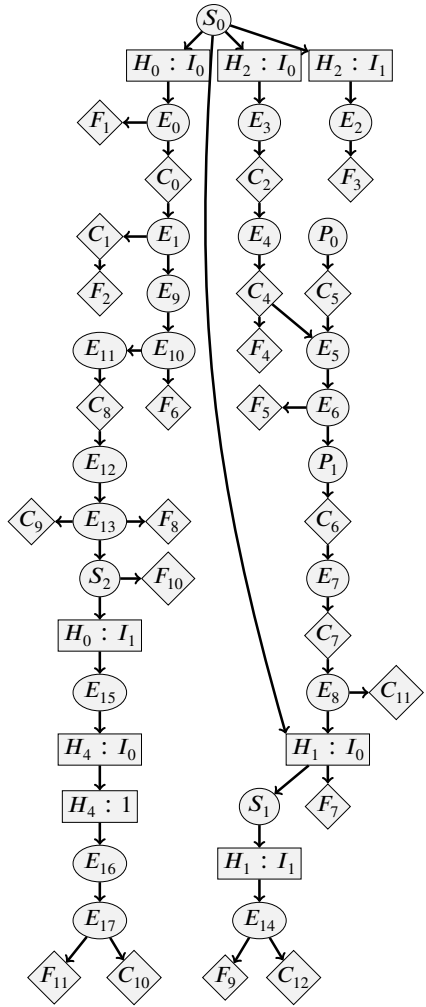
\begin{figure}[htb]

  \centering
\scalebox{0.6}{
\begin{tikzpicture}[scale=1,every text node part/.style={align=center,draw,font=\Large},  every edge/.style={draw=black, ultra thick}, node distance=4cm,auto, initial text = {}]

  \node (S0) [ellipse,inner sep=0.005cm, minimum width = 0.5cm,  fill=gray!10, draw] at (-3,2)  { {$S_{0}$}};

  \node (I2) [rectangle,  minimum width = 0.5cm,  fill=gray!10, draw] at (-4,1)  { {$H_{0}:I_{0}$}};
  \node (I3) [rectangle, minimum width = 0.5cm,  fill=gray!10, draw] at (-2,1)  { {$H_{2}:I_{0}$}};
  \node (I1) [rectangle, minimum width = 0.5cm,  fill=gray!10, draw] at (0,1)  { {$H_{2}:I_{1}$}};
  
  \node (T2) [diamond, inner sep=0.005cm, minimum width = 0.5cm,  fill=gray!10, draw] at (-5.5,-0.25)  { {$F_{1}$}};
  \node (E2) [ellipse, inner sep=0.05cm, minimum width = 0.5cm,  fill=gray!10, draw] at (-4,-0.25)  { {$E_{0}$}};
  \node (E3) [ellipse, inner sep=0.05cm, minimum width = 0.5cm,  fill=gray!10, draw] at (-2,-0.25)  { {$E_{3}$}};
  \node (E1) [ellipse, inner sep=0.05cm, minimum width = 0.5cm,  fill=gray!10, draw] at (0,-0.25)  { {$E_{2}$}};
  
  \node (T1) [diamond, inner sep=0.005cm, minimum width = 0.5cm,  fill=gray!10, draw] at (0,-1.5)  { {$F_{3}$}};
  
  \node (T3) [diamond, inner sep=0.005cm, minimum width = 0.5cm,  fill=gray!10, draw] at (-4,-1.5)  { {$C_{0}$}};
  \node (T4) [diamond, inner sep=0.005cm, minimum width = 0.5cm,  fill=gray!10, draw] at (-2,-1.5)  { {$C_{2}$}};
  \node (F1) [ellipse, inner sep=0.05cm, minimum width = 0.5cm,  fill=gray!10, draw] at (-4,-2.75)  { {$E_{1}$}};
  \node (F2) [ellipse, inner sep=0.05cm, minimum width = 0.5cm,  fill=gray!10, draw] at (-2,-2.75)  { {$E_{4}$}};
  \node (F3) [ellipse, inner sep=0.05cm, minimum width = 0.5cm,  fill=gray!10, draw] at (-0.5,-2.75)  { {$P_{0}$}};
  \node (G1) [diamond, inner sep=0.005cm, minimum width = 0.5cm,  fill=gray!10, draw] at (-5.5,-2.75)  { {$C_{1}$}};
  \node (G2) [ellipse, inner sep=0.05cm, minimum width = 0.5cm,  fill=gray!10, draw] at (-4,-4)  { {$E_{9}$}};
  \node (G3) [diamond, inner sep=0.005cm, minimum width = 0.5cm,  fill=gray!10, draw] at (-2,-4)  { {$C_{4}$}};
  \node (G4) [diamond, inner sep=0.005cm, minimum width = 0.5cm,  fill=gray!10, draw] at (-0.5,-4)  { {$C_{5}$}};
  \node (H1) [diamond, inner sep=0.005cm, minimum width = 0.5cm,  fill=gray!10, draw] at (-5.5,-4)  { {$F_{2}$}};
  \node (H2) [ellipse, inner sep=0.05cm, minimum width = 0.5cm,  fill=gray!10, draw] at (-4,-5.25)  { {$E_{10}$}};
  \node (H3) [diamond, inner sep=0.005cm, minimum width = 0.5cm,  fill=gray!10, draw] at (-2,-5.25)  { {$F_{4}$}};
  \node (H4) [ellipse, inner sep=0.05cm, minimum width = 0.5cm,  fill=gray!10, draw] at (-0.5,-5.25)  { {$E_{5}$}};
  \node (J1) [ellipse, inner sep=0.05cm, minimum width = 0.5cm,  fill=gray!10, draw] at (-5.5,-5.25)  { {$E_{11}$}};
  \node (J2) [diamond, inner sep=0.005cm, minimum width = 0.5cm,  fill=gray!10, draw] at (-4,-6.5)  { {$F_{6}$}};
  \node (J3) [ellipse, inner sep=0.05cm, minimum width = 0.5cm,  fill=gray!10, draw] at (-0.5,-6.5)  { {$E_{6}$}};
  \node (K0) [diamond, inner sep=0.005cm, minimum width = 0.5cm,  fill=gray!10, draw] at (-5.5,-6.5)  { {$C_{8}$}};
  \node (K1) [diamond, inner sep=0.005cm, minimum width = 0.5cm,  fill=gray!10, draw] at (-2,-6.5)  { {$F_{5}$}};
  \node (K2) [ellipse, inner sep=0.05cm, minimum width = 0.5cm,  fill=gray!10, draw] at (-0.5,-7.75)  { {$P_{1}$}};
  \node (L0) [ellipse, inner sep=0.05cm, minimum width = 0.5cm,  fill=gray!10, draw] at (-5.5,-7.75)  { {$E_{12}$}};
  \node (L1) [diamond, inner sep=0.005cm, minimum width = 0.5cm,  fill=gray!10, draw] at (-0.5,-9)  { {$C_{6}$}};

  \node (M0) [ellipse, inner sep=0.05cm, minimum width = 0.5cm,  fill=gray!10, draw] at (-5.5,-9)  { {$E_{13}$}};
  \node (N1) [diamond, inner sep=0.005cm, minimum width = 0.5cm,  fill=gray!10, draw] at (-4,-9)  { {$F_{8}$}};

  \node (N3) [diamond, inner sep=0.005cm, minimum width = 0.5cm,  fill=gray!10, draw] at (-7,-9)  { {$C_{9}$}};
  
  \node (M1) [ellipse, inner sep=0.05cm, minimum width = 0.5cm,  fill=gray!10, draw] at (-0.5,-10.25)  { {$E_{7}$}};
  \node (N0) [ellipse, inner sep=0.05cm, minimum width = 0.5cm,  fill=gray!10, draw] at (-5.5,-10.25)  { {$S_{2}$}};
  \node (O0) [diamond, inner sep=0.005cm, minimum width = 0.5cm,  fill=gray!10, draw] at (-4,-10.25)  { {$F_{10}$}};
  
  \node (N2) [diamond, inner sep=0.005cm, minimum width = 0.5cm,  fill=gray!10, draw] at (-0.5,-11.5)  { {$C_{7}$}};
  \node (I0) [rectangle, minimum width = 0.5cm,  fill=gray!10, draw] at (-5.5,-11.5)  { {$H_{0}:I_{1}$}};

  \node (O1) [ellipse, inner sep=0.05cm, minimum width = 0.5cm,  fill=gray!10, draw] at (-0.5,-12.75)  { {$E_{8}$}};

  \node (E0) [ellipse, inner sep=0.05cm, minimum width = 0.5cm,  fill=gray!10, draw] at (-5.5,-12.75)  { {$E_{15}$}};

  \node (I4) [rectangle, minimum width = 0.5cm,  fill=gray!10, draw] at (-0.5,-14)  { {$H_{1}:I_{0}$}};
  \node (T0) [rectangle,  minimum width = 0.5cm,  fill=gray!10, draw] at (-5.5,-14)  { {$H_{4}:I_{0}$}};

  \node (E4) [diamond, inner sep=0.005cm, minimum width = 0.5cm,  fill=gray!10, draw] at (-0.5,-15.25)  { {$F_{7}$}};
  
  \node (E5) [ellipse, inner sep=0.05cm, minimum width = 0.5cm,  fill=gray!10, draw] at (-2, -15.25)  { {$S_{1}$}};
  \node (F0) [rectangle, minimum width = 0.5cm,  fill=gray!10, draw] at (-5.5,-15.25)  { {$H_{4}:1$}};
  
  \node (I5) [rectangle, minimum width = 0.5cm,  fill=gray!10, draw] at (-2,-16.5)  { {$H_{1}:I_{1}$}};
  \node (G0) [ellipse, inner sep=0.05cm, minimum width = 0.5cm,  fill=gray!10, draw] at (-5.5,-16.5)  { {$E_{16}$}};
  
  \node (E6) [ellipse, inner sep=0.05cm, minimum width = 0.5cm,  fill=gray!10, draw] at (-2, -17.75)  { {$E_{14}$}};
  \node (H0) [ellipse, inner sep=0.05cm, minimum width = 0.5cm,  fill=gray!10, draw] at (-5.5,-17.75)  { {$E_{17}$}};
  
\node (T5) [diamond, inner sep=0.005cm, minimum width = 0.5cm,  fill=gray!10, draw] at (-2.75,-19)  { {$F_{9}$}};

\node (T6) [diamond, inner sep=0.005cm, minimum width = 0.5cm,  fill=gray!10, draw] at (-1.25,-19)  { {$C_{12}$}};
  
  \node (J0) [diamond, inner sep=0.005cm, minimum width = 0.5cm,  fill=gray!10, draw] at (-6.5,-19)  { {$F_{11}$}};
  
  \node (J7) [diamond, inner sep=0.005cm, minimum width = 0.5cm,  fill=gray!10, draw] at (-4.75,-19)  { {$C_{10}$}};

  \node (E7) [diamond, inner sep=0.005cm, minimum width = 0.5cm,  fill=gray!10, draw] at (1,-12.75)  { {$C_{11}$}};
  
  \path[->] 
  (S0) edge node {} (I1)
  (S0) edge node {} (I2)
  (S0) edge node {} (I3)
  (S0) edge[bend right=10] node {} (I4.north west)

  (I0) edge node {} (E0)
  (I1) edge node {} (E1)
  (I2) edge node {} (E2)
  (I3) edge node {} (E3)
  (I4) edge node {} (E4)
  (I4) edge node {} (E5)
  (I5) edge node {} (E6)
  (E5) edge node {} (I5)

  (E0) edge node {} (T0)
  (E1) edge node {} (T1)
  (E2) edge node {} (T2)
  (E2) edge node {} (T3)
  (E3) edge node {} (T4)
  (E6) edge node {} (T5)
  (E6) edge node {} (T6)

  (T0) edge node {} (F0)
  (T3) edge node {} (F1)
  (T4) edge node {} (F2)
  
  (F0) edge node {} (G0)
  (F1) edge node {} (G1)
  (F1) edge node {} (G2)
  (F2) edge node {} (G3)
  (F3) edge node {} (G4)

  (G0) edge node {} (H0)
  (G1) edge node {} (H1)
  (G2) edge node {} (H2)
  (G3) edge node {} (H3)
  (G3) edge node {} (H4)
  (G4) edge node {} (H4)

  (H0) edge node {} (J0)
  (H2) edge node {} (J1)
  (H2) edge node {} (J2)  
  (H4) edge node {} (J3)
  (H0) edge node {} (J7)

  (J1) edge node {} (K0)
  (J3) edge node {} (K1)
  (J3) edge node {} (K2)

  (K0) edge node {} (L0)
  (K2) edge node {} (L1)

  (L0) edge node {} (M0)
  (L1) edge node {} (M1)

  (M0) edge node {} (N0)
  (M0) edge node {} (N1)
  (M0) edge node {} (N3)
  (M1) edge node {} (N2)

  (N0) edge node {} (O0)
  (N0) edge node {} (I0)
  (N2) edge node {} (O1)

  (O1) edge[] node {} (I4)
  (O1) edge node {} (E7)
  ;
    
\end{tikzpicture}
}
  
  \caption[justification=centering]{Lab environment exploitation graph.}
  \label{Fig:Lab.Environment.Graph}
\end{figure}

\subsubsection*{ADAPT Statistics}

  Fig.~\ref{Fig:Lab.RunningTime} shows the running time of each component of ADAPT for each of our case studies with the concurrent target and attack thresholds set to 1.
In all three cases, the vast majority of the running time is spent on activities of the managed system, i.e., in the operation of scanning, exploitation and post-exploitation tools. In our experiments, only a few of post-exploitation tools were required: 1 for Metasploitable2 and Metasploitable3 and 2 for the VM network. In a different case study, more such operations might be required. The MAPE-K loop operations take a few tens of miliseconds at most.

\begin{figure}[htb]
\begin{subfigure}[b]{1\linewidth}
\centering
\scalebox{0.9}{
\begin{tikzpicture}[]
    \begin{axis}[
        width  = 1.1\linewidth,
        height = 5cm,
        major x tick style = transparent,
        ybar=\pgflinewidth,
        bar width=5pt,
        ymajorgrids = true,
        ylabel = {Running Time(ms)},
        symbolic x coords={Metasploitable2,Metasploitable3, VM Network},
        xtick = data,
        scaled y ticks = false,
        enlarge x limits=0.25,
        ymin=0,
        legend style={at={(0.5,1.2)},cells={align=center},
          anchor=north,legend columns=4},
      ]
       
      \addplot[style={black,fill=blue,mark=none}]
      coordinates {(Metasploitable2, 1) (Metasploitable3,1) (VM Network,3)};
    
      \addplot[style={black,fill=yellow,mark=none},opacity=0.7]
      coordinates {(Metasploitable2, 1) (Metasploitable3,2) (VM Network,2)};
      
      \addplot[style={black,fill=red,mark=none}]
      coordinates {(Metasploitable2, 2) (Metasploitable3,4) (VM Network,12)};
      
      \addplot+[style={black,fill=teal,mark=none},opacity=1]
      coordinates {(Metasploitable2, 1) (Metasploitable3,1) (VM Network,5)};  
    \legend{Monitor, Analyze, Plan, Execute}
    \end{axis} 
\end{tikzpicture}
}
\caption{MAPE Components}
\end{subfigure}

\begin{subfigure}[b]{\linewidth}
\centering
\scalebox{0.9}{
\begin{tikzpicture}[]
    \begin{axis}[
        width  = 1.1\linewidth,
        height = 5cm,
        major x tick style = transparent,
        ybar=\pgflinewidth,
        bar width=5pt,
        ymajorgrids = true,
        ylabel = {Running Time(ms)},
        symbolic x coords={Metasploitable2,Metasploitable3, VM Network},
        xtick = data,
        scaled y ticks = false,
        enlarge x limits=0.25,
        ymin=0,
        legend style={at={(0.5,1.20)},cells={align=center},
          anchor=north,legend columns=4},
      ]
      \addplot[style={black,fill=Thistle,mark=none}]
      coordinates {(Metasploitable2, 59000) (Metasploitable3, 121000) (VM Network,129500)};
      
	  \addplot+[style={black,fill=Goldenrod,mark=none},opacity=1, ]
      coordinates {(Metasploitable2,97000) (Metasploitable3,142500) (VM Network,228410)};
      
	  \addplot+[style={black,fill=RoyalPurple,mark=none},opacity=1, ]
      coordinates {(Metasploitable2,100) (Metasploitable3,1000) (VM Network,21000)};

      \addplot+[style={black,fill=Periwinkle,mark=none},opacity=1, ]
      coordinates {(Metasploitable2,157000) (Metasploitable3,265500) (VM Network,387000)};
        
    \legend{Scan, Exploit, Post-Exploit, Total}
    \end{axis}
\end{tikzpicture}
}
\caption{Managed System Components}
\end{subfigure}
 \caption[justification=centering]{ Running time per ADAPT component}
  \label{Fig:Lab.RunningTime}
\end{figure}
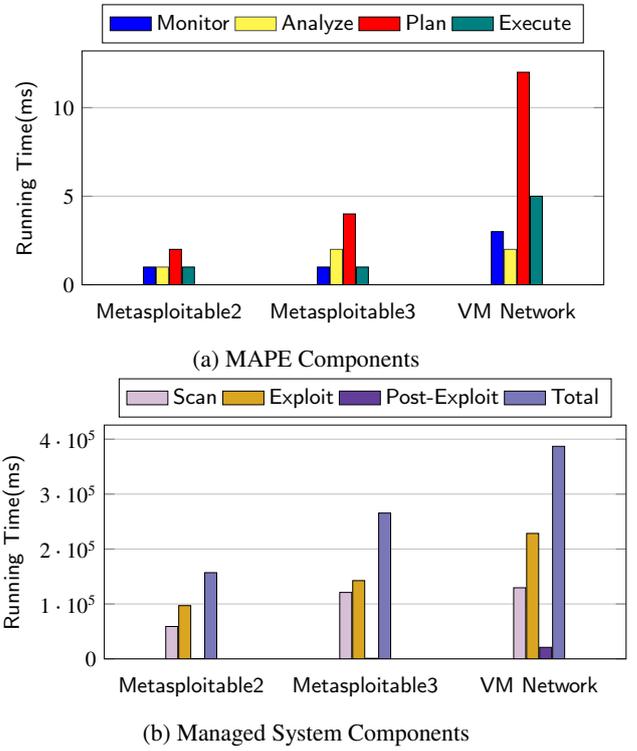

Figure~\ref{Fig:Lab.Utility} shows the cumulative utility score after each adaptation step for different target and attack thresholds in the VM network case study.
As expected, since the same repertoires are being used in all three runs against the VM network, utility converges to the same final score.
The differences in utility per adaptation step, can be attributed to the different paths taken by the tool depending on the amount of concurrency allowed.
The corresponding adaptation step running times are given in Figure~\ref{Fig:Lab.AdaptationTime}, predictably, as the concurrency thresholds increase, the total running time decreases.
Increasing the thresholds from 1 to 2, noticeably improves the running time, however a further threshold increase from 2 to 3 offers minimal improvement.
We expect this to be dependent on the amount of independent exploitation paths in the target system. Indeed, by contrasting
with Figure~\ref{Fig:Lab.Environment.Graph}, we can see that 
there are mainly two independent exploitation paths present in the
VM network.
A noteworthy observation is that fewer adaptation steps do not necessarily imply a lower total running time. This fact can be attributed to two factors: (i) the fact that the MAPE-K loop operation
takes a negligible amount of time, and (ii) the event-based nature of ADAPT's implementation, which allows adaptations to be triggered as soon as any scan, exploitation step or post-exploitation tool run has completed. 

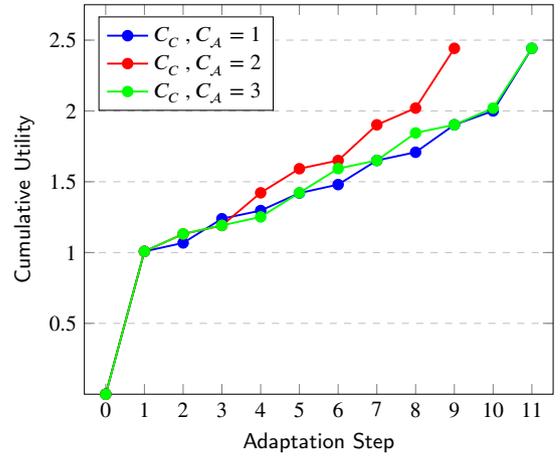
\begin{figure}[htb]
\centering
\scalebox{0.9}{
\begin{tikzpicture}[]
\begin{axis}[
    xlabel={Adaptation Step},
    ylabel={Cumulative Utility},
    xmin=0, xmax=11,
    ymin=0, ymax=2.75,
    xtick={0,1,2,3,4,5,6,7,8,9,10,11},
    ytick={0.5, 1, 1.5, 2,2.5},
    legend pos=north west,
    ymajorgrids=true,
	enlarge x limits=0.05,
    grid style=dashed,
    every axis plot/.append style={thick}
]
 
\addplot[
    color=blue,
    mark=*,
    ]
coordinates {
(0,  0      )
(1, 1.0082	)
(2, 1.0682	)
(3, 1.2388	)
(4, 1.2964	)
(5, 1.4200	)
(6, 1.4800	)
(7, 1.6505	)
(8, 1.7082	)
(9, 1.9023	)
(10, 2.0	)
(11, 2.442	)
  
};

\addplot[
    color=red,
    mark=*,
    ]
coordinates {
(0, 0)
(1, 1.008)
(2, 1.131)
(3, 1.191)
(4, 1.422)
(5, 1.592)
(6, 1.650)
(7, 1.902)
(8, 2.02 )
(9, 2.442)
};

\addplot[
    color=green,
    mark=*,
    ]
coordinates {

(0, 0)
(1, 1.0082)
(2, 1.1317)
(3, 1.1917)
(4, 1.2517)
(5, 1.4223)
(6, 1.5929)
(7, 1.6505)
(8, 1.8447)
(9, 1.9023)
(10, 2.02	)
(11, 2.442)

};

\legend{$C_{C} \mathbin{,} C_{\mathcal{A}}=1$, $C_{C} \mathbin{,} C_{\mathcal{A}}=2$, $C_{C} \mathbin{,} C_{\mathcal{A}}=3$}
\end{axis}
\end{tikzpicture}
}
 \caption[justification=centering]{VM network cumulative utility}
  \label{Fig:Lab.Utility}
\end{figure}
\begin{figure}[h]
\centering
\scalebox{0.9}{
\begin{tikzpicture}[]
\begin{axis}[
    xlabel={Adaptation Step},
    ylabel={Time (s)},
    xmin=0, xmax=11,
    ymin=0, ymax=400,
    xtick={0,1,2,3,4,5,6,7,8,9,10,11},
    ytick={25, 50,75,100,125,150,175,200,225,250,275,300,325,350,375,400},
    legend pos=north west,
    ymajorgrids=true,
	enlarge x limits=0.05,
    grid style=dashed,
    every axis plot/.append style={thick}
]
 
\addplot[
    color=blue,
    mark=*,
    ]
coordinates {
(0,  0)
(1,  30)
(2,  33)
(3,  49)
(4,  111)
(5,  121)
(6,  123)
(7,  125)
(8,  196)
(9,  207)
(10, 386)
(11, 387)

};

\addplot[
    color=red,
    mark=*,
    ]
coordinates {
(0, 0)
(1, 29)
(2, 30)
(3, 31)
(4, 33)
(5, 49)
(6, 90)
(7, 100)
(8, 278)
(9, 279)
};

\addplot[
    color=green,
    mark=*,
    ]
coordinates {
  (0, 0)
  (1, 22)
  (2, 29)
  (3, 30)
  (4, 31)
  (5, 33)
  (6, 39)
  (7, 81)
  (8, 92)	
  (9, 104)	
  (10, 270)
  (11, 271)

};

\legend{$C_{C} \mathbin{,} C_{\mathcal{A}}=1$, $C_{C} \mathbin{,} C_{\mathcal{A}}=2$, $C_{C} \mathbin{,} C_{\mathcal{A}}=3$}
\end{axis}
\end{tikzpicture}
}
 \caption[justification=centering]{VM network time per adaptation step.}
  \label{Fig:Lab.AdaptationTime}
\end{figure}
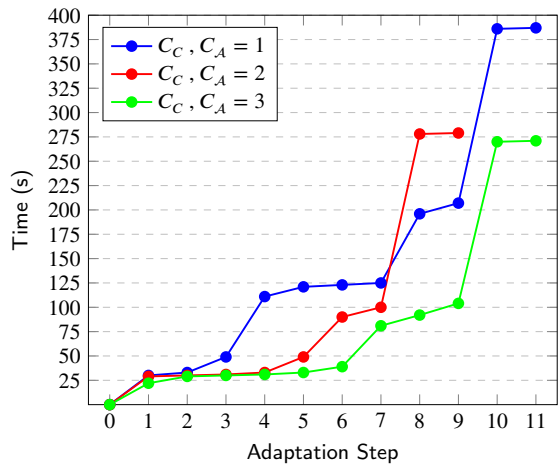


\section{Conclusion}
\label{sec::conclusion}

We have formulated penetration testing at the architectural level and proposed a general self-adaptive architecture to automate penetration testing. We have further described the design and implementation of  \toolname, a tool that can automate the penetration testing of host and service-style systems.
Our evaluation results indicate that \toolname is capable 
of automating the execution of penetration tests against realistic 
systems. To our knowledge, \toolname is the fist tool that is able to automate all parts of the penetration testing process including
planning, dynamic at runtime decision making and attack automation that can be employed against realistic environments.

As future work, we plan to work on expanding the autonomy of our approach. A promising direction is towards incorporating a method for automated repertoire generation, which involves the capability to automatically synthesize attack tactics from constituent attack techniques. Moreover we plan to investigate the effectiveness of the proposed architecture in different domains and as a component of a larger cybersecurity tool ecosystem.

\subsection*{Statement on Source and Data Availability}
We make the tool source code available at: \url{\coderepo}.
Please note that \toolname is plugin-based and due to ethical considerations, the \emph{intrusive} or operationally \emph{disruptive}
plugins will only be made available to researchers upon request. 
In terms of data availability of our case studies, Metasploitable2 and Metasploitable3 are open and available freely online. When it comes to the VM network, we can not openly distribute it since it is being used as examination material in active courses but we can provide access upon request.

\bibliographystyle{splncs04}
\bibliography{bib.bib}
\appendix
\appendix
\clearpage

\section{Appendix A}

\begin{table}[!htbp]
   \caption{Target Decision Utility}
  \label{tab:target.properties.full}
  \footnotesize
  \centering
    \let\Tabular\tabular
    \def\tabular{\Tabular}
   \begin{tabular}{ccc}
      \toprule
    \textbf{Factor} & \textbf{Weight} & \textbf{Value} \\
      \midrule
       Number of Services(S) & 0.2 & \large{$\frac{\textup{min}(\textup{S},10)}{10}$} \\
       \midrule
       Number of Vulnerabilities(V) & 0.2 & \large{$\frac{\textup{min}(\textup{V}^{2},10)}{10}$} \\
       \midrule
       Number of Connections(C) & 0.2 &  \large{$\frac{\textup{min}(\textup{C}^{1.5},10)}{10}$} \\
        \midrule
      Exploitation State  (E) & 0.4 & 
$ \begin{cases}
                    0.73, &   E = None  \\
                    0.9,  &   E = Initial\\
                    1.0,  &   E = Elevated \\
                    0.23, &   E = C2C
                 \end{cases} $\\
                 \bottomrule
    \end{tabular}
\end{table}

\begin{table}[!htbp]
\caption{Attack Decision Utility}
\label{tab:attack.properties.full}
\footnotesize
\centering
\begin{tabular}{ccc}
\toprule
\textbf{Factor} & \textbf{Weight} & \textbf{Value}  \\
\midrule
Attack Vector (V) & 0.3 & 
$ \begin{cases}
                    1.0, &   V = Network  \\
                    0.73, &  V = Adjacent\\
                    0.64,  &  V = Local \\
                    0.23, &   V = Physical
                 \end{cases} $\\
\hline
Attack Complexity (C) & 0.2 & 
$ \begin{cases}
                    1.0, &   C = Low  \\
                    0.57,  &  C = High\\
                 \end{cases} $\\
\hline
Required Privileges (P) & 0.2 & 
$ \begin{cases}
                    1.0, &   P = None  \\
                    0.73,  &  P = Low\\
                    0.32 &   P = High\\
                 \end{cases} $\\
\hline
User Interaction (U)  & 0.1 & 
$ \begin{cases}
                    1.0, &   U = None  \\
                    0.73,  &  U = Required\\
                 \end{cases} $\\
\hline
Running Time  (T) & 0.2 & 
$ \begin{cases}
                    1.0, &   T = Quick  \\
                    0.73,  &  T = Medium\\
                    0.31,  &  T = Long \\
                    0.12, &   T = Guessing
                 \end{cases} $\\
\bottomrule
\end{tabular}
\end{table}

\begin{table}[!htbp]
\caption{Scan Decision Utility}
\label{tab:scan.properties.full}
\footnotesize
\centering
\begin{tabular}{ccc}
\toprule
\textbf{Factor} & \textbf{Weight} & \textbf{Value} \\
\midrule
Scan Range(R) & 0.25 & 
$ \begin{cases}
                    1.0, &   R = Network  \\
                    0.73,  & R = Host\\
                    0.32,  & R = Local \\
                 \end{cases} $\\

\hline
Number of Targets (T) & 0.25 &
$ \begin{cases}
                    1.0, &   T = Network  \\
                    0.73,  &  T = Multiple\\
                    0.32,  &  T = One 
                 \end{cases} $\\
\hline
Scan Duration(D) & 0.25 & 
$ \begin{cases}
                    1.0, &   D = Quick  \\
                    0.73,  &  D = Medium\\
                    0.31,  &  D = Long \\
                    0.12, &   D = Guessing
                 \end{cases} $\\
\hline
Scan Complexity(C) & 0.25 & 
$ \begin{cases}
                    1.0, &   C = Low  \\
                    0.73,  &  C = High
                 \end{cases} $\\
\bottomrule
\end{tabular}
\end{table}

\subsection*{Mapping ~\cite{Zhou2019} to our model}
  Since ~\cite{Zhou2019} is the closest related work that uses reinforcement-learning to achieve black-box penetration testing,
  we show how to transform its  representation of the penetration testing problem to our formalisms.
  Other related work that utilizes reinforcement learning, e.g., ~\cite{9394285,9229752,app11198823} can be transformed following a similar process to the one we describe below.
  
\begin{example}
  In \cite{Zhou2019}, a penetration test is given by an markov decision process~\cite{puterman2014markov}, $\mathcal{M} = \langle \Phi, \Psi, \mathcal{P}_{\Psi}, \mathcal{R}_{\Psi} \rangle$, where:
  \begin{itemize}
  \item $\phi \in \Phi$ is a tuple $\langle \iota, o, \theta, \upsilon, \epsilon \rangle$, that corresponds to a host machine. $\iota$ is a machine identifier, $o$ corresponds to the machine's operating system, $\theta$ is the set of the machine's open ports, $\upsilon$ is the set of known services running on the machine and $\varepsilon$ is the machine's exploitation state,
  \item    $\Psi$ is the set of available penetration test actions,  $\mathcal{P}_{\Psi}$, is the probability that an action $\psi \in \Psi$ will lead from a state $\phi$ to another state $\phi^{'}$, and $\mathcal{R}_{\Psi}$, is the corresponding reward.
  \end{itemize}

The following transformation maps $\mathcal{M}$  to $T$:
    \begin{enumerate}
    \item Every $\phi \in \Phi$ is mapped to a component $c_{\phi} \in C$, $\iota$, $o$ and
      $\varepsilon$ are each mapped to corresponding component properties of $c_{\phi}$, $\pi_{\iota} \in \Pi_{\phi}$, $\pi_{o} \in \Pi_{\phi}$ and $\pi_{\varepsilon} \in \Pi_{\phi}$. For each service, $\upsilon$ we create an interface $i_{\upsilon} \in I_{\phi}$ and the corresponding open port $\theta$ is a property: $\pi_{\theta} \in \Pi_{\upsilon}$.
    \item Every action $\psi \in \Psi$ is mapped to either a scan $\sigma \in \Sigma$ or to an attack step $\alpha$. Each attack step also constitutes an attack $A \in \Attacks$, since the authors in ~\cite{Zhou2019} make no distinction between attack steps and complete attacks.
      \item Action probabilities $\mathcal{P}_{\Psi}$ and rewards, $\mathcal{R}_{\Psi}$ can be modeled through a strategy $S^{\mathcal{M}} = \langle M^{\mathcal{M}}, M^{\mathcal{M}}_{0}, M^{\mathcal{M}}_{+}, N^{\mathcal{M}}_{+} \rangle$ where $M^{\mathcal{M}}$ stores $\mathcal{P}_{\Psi}$ and $\mathcal{R}_{\Psi}$, $M^{\mathcal{M}}_{+}$ is used to update their values and $N^{\mathcal{M}}_{+}$ returns the action $\psi$ with the highest reward in the current state $\phi$. 
      \end{enumerate}
\end{example}

\subsection*{Implementation and Case Study Details}
\label{app:implementation.details}
\begin{table}[htbp]
  \caption{Example Managed system tools and their classification}
  \label{tab:tools}
  \centering
    \let\Tabular\tabular
    \def\tabular{\Tabular}
    \scalebox{0.7}{
    \begin{tabular}{ccc}
      \toprule
       \textbf{Tool Name} & \textbf{Tool Role} & \textbf{ATT\&CK Technique} \\
      \midrule
      nmap & probe & Active Scanning: Scanning IP Blocks  \\
      skipfish & probe & Active Scanning: Wordlist Scanning \\
      NSE discovery & probe  &  Network Service Discovery \\
      custom reporter  & probe  & Automated Exfiltration \\
      \midrule
    sqlmap & exploitation & Exploit Public-Facing Application \\
      NSE brute & exploitation  & Brute Force: Password Cracking  \\
     hydra & exploitation & Brute Force: Password Cracking\\
     msfvenom & exploitation & Develop Capabilities: Exploits  \\
     metasploit  & exploitation  & Exploitation for Privilege Escalation\\
    cron shell & exploitation  & Scheduled Task/Job: Cron\\
    pickle shell generator  & exploitation  & Exploitation for Privilege Escalation\\
    \midrule
     mimikatz  & post-exploitation & OS Credential Dumping \\
     tcpdump  & post-exploitation & Network Sniffing \\
     linuxprivchecker  & post-exploitation  & Log Enumeration \\
     hashcat  & post-exploitation  & Brute Force: Password Cracking \\
     netcat  & post-exploitation  & Remote  \\
     curl  & post-exploitation  & Network Swiss Knife  \\
     netstat  & post-exploitation  & Network Service Discovery \\
     ps  & post-exploitation  & Process Discovery \\
     \bottomrule
    \end{tabular}
}
\end{table}

\label{app::case.studies}
    \begin{table}[htbp]
  \caption{Flag descriptions}
  \label{tab:flag.descriptions}
  \footnotesize
  \centering
    \let\Tabular\tabular
    \def\tabular{\Tabular}
    \scalebox{0.7}{
    \begin{tabular}{|c|p{38em}|}
      \hline
      \textbf{ID} & \textbf{Description}  \\
      \hline
      0 & Perform reconnaissance and discovery to plan operations  \\
      \hline
      1 & \makecell[l]{ (i) Execute malicious code on remote devices
     \\ (ii) Collect and exfiltrate data from computing environments } \\
      \hline
      2 & \makecell[l]{Perform reconnaissance and discovery to plan operations access credentials, such as:\\ account names, passwords and access tokens } \\
      \hline
      3 & \makecell[l]{(i)Perform reconnaissance and discovery to plan operations
 access credentials, such as: \\ account names, passwords and access tokens
\\(ii) Achieve initial access to networks and systems  }\\
    \hline
     4 & \makecell[l]{ Perform reconnaissance and discovery to: \\
     (i) plan operations
\\(ii) achieve initial access to networks and systems
\\(iii) execute malicious code on remote devices
\\(iv) establish command and control capabilities to communicate with compromised systems
\\(v) persist on networks by maintaining access across interruptions
\\(vi) move laterally, pivoting through the computing environment   }\\
     \hline
     5 & \makecell[l]{(i) Execute malicious code on remote devices
\\(ii) Achieve initial access to networks and systems
\\(iii) Establish command and control capabilities to communicate with compromised systems.   }\\
      \hline
     6 &  \makecell[l]{(i) Perform reconnaissance and discovery to plan operations
\\ (ii) Access credentials, such as account names, passwords and access tokens
\\ (iii) Achieve initial access to networks and systems
\\ (iv) Execute malicious code on remote devices
\\ (v) establish command and control capabilities to communicate with compromised systems
\\ (vi) Move laterally, pivoting through the computing environment
\\ (vii) Collect and exfiltrate data from computing environments. } \\
     \hline
     7 &  \makecell[l]{ (i) Collect and exfiltrate data from computing environments
\\  (ii) Execute malicious code on remote devices} \\
      \hline
     8 & \makecell[l]{ (i) Execute malicious code on remote devices
\\ (ii) Establish command and control capabilities to communicate with compromised systems
\\ (iii) elevate privileges on systems to gain higher-level permissions
\\ (iv) collect and exfiltrate data from computing environments }\\
     \hline
     9 & \makecell[l]{ (i) Perform reconnaissance and discovery to plan operations
\\  (ii) collect and exfiltrate data from computing environments  }\\
      \hline
      10 & \makecell[l]{ (i) Establish resources to support offensive security operations
\\ (ii) Achieve initial access to networks and systems
\\ (iii) Execute malicious code on remote devices
\\ (iv) Establish command and control capabilities to communicate with compromised systems
\\ (v) Persist on networks by maintaining access across interruptions
\\ (vi) Move laterally, pivoting through the computing environment   } \\
      \hline
    \end{tabular}
    }
\end{table}

\label{Exploitation.Graph.Descriptions}
\begin{table}[ht]
\caption{Metasploitable2 Exploitation Graph Details}
\label{tab:metasploitable2.detail}
\footnotesize
\centering
    \let\Tabular\tabular
    \def\tabular{\Tabular}
    \scalebox{0.75}{
   \begin{tabular}{|c|c|c|}
      \hline
    \textbf{Node} & \textbf{Description}  & \textbf{\makecell{Running \\ Time (s)}} \\
      \hline
       S0  & NetworkScan & 30 \\
      \hline
      S1  & LinPEAS & 7 \\
      \hline
      S2  & LinEnum.sh  & 9  \\
      \hline
      S3  & linuxprivchecker.py  & 12 \\
      \hline
       E0   & SSH Bruteforce & 10 \\
      \hline
       E1   & Credential Leak & 2 \\
      \hline      
      E2   & User Enumeration & 7\\
      \hline      
      E3   & Password Bruteforce &  5.8\\
      \hline      
      E4  & CGI Argument Injection & 8.8 \\
      \hline      
      E5  & Backdoor & 8 \\
      \hline      
      E6  & Credential Bruteforce & 5.69\\
      \hline      
      E7  & Unauthenticated Access & 1\\
      \hline      
      E8  & Unauthenticated Access & 1\\
      \hline      
      E9  & Unauthenticated Access & 1\\
      \hline      
      E10  & Directory Mount & 2\\
      \hline      
      E11  & Backdoor & 17\\
      \hline      
      E12  & Command Execution & 3\\
      \hline      
      E13  & Remote Class Loading & 16\\
      \hline      
      E14  & Remote Shell & 5\\
      \hline
      E15  & Remote Shell & 1\\
      \hline
      E16 &  CVE-2008-0900 & 1\\
      \hline   
      E17  & CVE-2008-0600 & 1\\
      \hline    
      E18  & CVE-2009-1046 & 1\\
      \hline    
      P0  & Passwordless sudo & 0.1\\
      \hline    
    \end{tabular}
    \quad
    \begin{tabular}{|c|c|}
      \hline
    \textbf{Node} & \textbf{Description} \\
      \hline
       I0 & ssh \\
      \hline
       I1 & telnet  \\
      \hline
       I2 & smtp  \\
      \hline      
      I3 & postgress  \\
      \hline      
      I4 & httpd  \\
      \hline      
      I5 & vsftpd  \\
      \hline      
      I6 & vnc   \\
      \hline      
      I7 & rlogin \\
      \hline      
      I8 & rsh \\
      \hline      
      I9 & rexec  \\
      \hline      
      I10 & NFS \\
      \hline      
      I11 & UnrealIRCD \\
      \hline      
      I12 & samba  \\
      \hline      
      I13 & javarmi   \\
      \hline      
      I14 & mysql  \\
      \hline      
      C1 & CVE-2008-0900 exploitable \\
      \hline      
      C2 & CVE-2008-0600 exploitable  \\
      \hline
      C3 & CVE-2009-1046 exploitable \\
      \hline    
      C4 & msfadmin credentials \\
      \hline    
      C5 & www-data shell \\
      \hline
      C6 & root credentials \\
      \hline
    \end{tabular}
    }
    \end{table}

    \begin{table}[ht]
\caption{Metasploitable3 Exploitation Node Details}
\label{tab:metasplotable3.detail}
\footnotesize
\centering
    \let\Tabular\tabular
    \def\tabular{\Tabular}
    \scalebox{0.75}{
   \begin{tabular}{|c|c|c|}
      \hline
    \textbf{Node} & \textbf{Description} & \textbf{\makecell{Running \\ Time}} \\
      \hline
       S0 & NetworkScan & 120  \\
       \hline
       S1 & EternalBlue Scan & 1 \\
      \hline
       E0 & Credential Bruteforce & 9 \\
      \hline
       E1 & Credential Bruteforce & 8\\
      \hline      
      E2 & Credential Bruteforce & 7 \\
      \hline      
      E3 & Credential Bruteforce & 8\\
      \hline      
      E4 & OS command execution & 1 \\
      \hline      
      E5 & Credential Bruteforce & 7\\
      \hline      
      E6 & Remote Code Execution & 1\\
      \hline      
      E7 & Upload Remote Shell & 2 \\
      \hline      
      E8 & Upload Remote Shell & 2\\
      \hline      
      E9 & Remote Class Load &  2 \\
      \hline      
      E10 & Remote Code Execution & 1.5 \\
      \hline      
      E11 & Credential Bruteforce & 8\\
      \hline      
      E12 & Credential Bruteforce & 9\\
      \hline      
      E13 & Credential Bruteforce & 12\\
      \hline      
      E14 & Credential Bruteforce & 14\\
      \hline
      E15 & Credential Bruteforce & 11\\
      \hline
      E16 & Credential Bruteforce & 8\\
      \hline
      E17 & Credential Bruteforce & 15 \\
      \hline
      E19 & TCP Shell War Upload & 2\\
      \hline
      E20 & TCP Shell War Upload & 2 \\
      \hline
      E21 & Backdoor Access &1  \\
      \hline
      E22 & Remote Code Execution & 1  \\
      \hline
      E23 & Upload Remote Shell & 2  \\
      \hline
      E24 & Upload Remote Shell & 2 \\
      \hline
      E26 & shell access & 1 \\
      \hline
      E27 & shell access & 1 \\
      \hline
      E28 & Reverse TCP shell & 1 \\
      \hline
      E29 & Reverse TCP root shell & 1 \\
      \hline
      E30 & Eternal Blue & 1 \\
      \hline
      E31 & CVE2015-1635 & 1 \\
      \hline
      P0 & Malicious War Payload & 1 \\
      \hline    
    \end{tabular}
    \begin{tabular}{|c|c|}
      \hline
    \textbf{Node} & \textbf{Description} \\
      \hline
       I0 & ssh  \\
      \hline
       I1 & winrm \\
      \hline
       I2 & tomcat \\
      \hline      
      I3 & glassfish  \\
      \hline      
      I4 & jenkins  \\
      \hline      
      I5 & caidao  \\
      \hline      
      I6 & elasticsearch  \\
      \hline      
      I7 & axis  \\
      \hline      
      I9 & webdav  \\\hline      
      I10 & jmx  \\
      \hline      
      I11 & wordpress  \\\hline      
      I12 & PHPMyAdmin  \\
      \hline      
      I13 & rubyonrails  \\
      \hline      
      I14 & struts  \\
      \hline      
      I14 & manageengine  \\
      \hline      
      I15 & iis-ftp  \\
      \hline      
      I16 & mysql  \\
      \hline      
      I17 & iis-http  \\
      \hline      
      C0 & vagrant credentials  \\
      \hline      
      C1 & vagrant credentials  \\
      \hline      
      C2 & vagrant credentials  \\
      \hline
      C3 & splot credentials  \\
      \hline    
      C4 & vagrant credentials  \\
      \hline    
      C5 & root credentials  \\
      \hline    
      C6 & sploit credentials  \\
      \hline    
      C7 & admin credentials  \\
      \hline    
      C8 & directory traversal  \\
      \hline    
      C9 & root credentials  \\
      \hline
      C10 & root credentials  \\
      \hline
      C11 & root credentials  \\
      \hline
      C12 & NT Authority System creds  \\
      \hline
      C13 & Process memory dump  \\
      \hline
    \end{tabular}
    }
\end{table}

\end{document}